\documentclass[journal]{IEEEtran}
\IEEEoverridecommandlockouts
\usepackage{pgfplots}
\pgfplotsset{compat=1.18}
\usepackage{xcolor}
\usepackage{pifont}
 \usepackage{url}
 \usepackage[most]{tcolorbox}
\usepackage[
  colorlinks=false,
  linkbordercolor={0 1 0},   % RGB values for green
  citebordercolor={0 1 0},
  urlbordercolor={0 1 0}
]{hyperref}

\newtcolorbox{rqbox1}{
    colback=gray!8,
    colframe=black,
    boxrule=0.8pt,
    arc=2pt,
    left=6pt,
    right=6pt,
    top=6pt,
    bottom=6pt,
    title=System Prompt,
    fonttitle=\bfseries
}
\definecolor{GOne}{RGB}{33,113,181}     % black
\definecolor{GTwo}{RGB}{35,139,69}      % Green
\definecolor{GThree}{RGB}{146,36,40}    % Deep red
\definecolor{GFour}{RGB}{117,107,177}   % Muted purple
\definecolor{GFive}{RGB}{217,95,2}      % Dark orange
\definecolor{EdgeColor}{RGB}{0,90,120}      % Deep teal
\definecolor{EdgeColorLight}{RGB}{0,130,160}
\usepackage{booktabs}
\usepackage{cite}
\usepackage{amsmath,amssymb,amsfonts}
\usepackage{algorithmic}
\usepackage{multirow}
\usepackage{graphicx}
\usepackage{textcomp}
\usepackage{xcolor}
\def\BibTeX{{\rm B\kern-.05em{\sc i\kern-.025em b}\kern-.08em
    T\kern-.1667em\lower.7ex\hbox{E}\kern-.125emX}}
\begin{document}
\bstctlcite{BSTcontrol}

\title{6G Needs Agents: Toward Agentic AI-Native Networks for Autonomous Intelligence}

\author{
\IEEEauthorblockN{
Mohamed~Amine~Ferrag\IEEEauthorrefmark{1}\IEEEauthorrefmark{3},
Abderrahmane~Lakas\IEEEauthorrefmark{1},
Merouane~Debbah\IEEEauthorrefmark{2}
}
\\
\IEEEauthorblockA{\IEEEauthorrefmark{1}
Department of Computer and Network Engineering,  
United Arab Emirates University, UAE
} \\
\IEEEauthorblockA{\IEEEauthorrefmark{2}
Research Institute for Digital Future, Khalifa University, UAE
}\\
\IEEEauthorblockA{\IEEEauthorrefmark{3}
Corresponding author: \texttt{mohamed.ferrag@uaeu.ac.ae}
}
}

\maketitle

\begin{abstract}
Sixth-generation (6G) networks are increasingly envisioned as AI-native infrastructures integrating communication, sensing, and computing into a unified fabric. However, existing approaches remain largely optimization-centric, relying on closed-loop control with limited reasoning capability. In this paper, we argue for a paradigm shift toward \emph{Agentic AI-Native 6G}, in which Large Language Model (LLM)-based agents operate as bounded, policy-governed reasoning entities within a semantic control plane layered above deterministic 3GPP infrastructure. We propose a four-layer architecture that integrates deterministic network infrastructure, semantic abstraction of intent and context, hierarchical reasoning, and a distributed multi-agent fabric spanning device, edge, and core domains. To assess feasibility, we develop a proof-of-concept agentic reasoning and orchestration framework and conduct an extensive empirical study using a domain-specific 6G benchmark under realistic deployment constraints. Our results reveal a fundamental tradeoff between reasoning capability and system efficiency, showing that no single model simultaneously satisfies latency, throughput, and accuracy requirements. Instead, heterogeneous deployment of LLM agents across the device--edge--core continuum is necessary to balance these constraints. We further demonstrate that quantization introduces non-uniform effects across models, reinforcing the need for system-level optimization rather than model-level compression alone. These findings establish agentic intelligence as a viable architectural direction for 6G and highlight key challenges in achieving scalable, trustworthy, and self-reasoning networks. All experimental results and evaluation scripts are publicly available to support reproducibility at \url{https://github.com/maferrag/6G-Bench/tree/main/Agentic6G}.
\end{abstract}


\section{Introduction}

Sixth-generation (6G) networks are transitioning from a conceptual vision to a structured, globally coordinated standardization process \cite{wu2026toward,jiang2026agentic,luo2026trustworthy}. Within 3GPP, Release 20 marks a pivotal inflection point, advancing 5G-Advanced specifications while initiating formal 6G study items across the Service and System Aspects (SA), Radio Access Network (RAN), and Core Network (CT) groups. As outlined in the Release 20 work plan, Stage 1 (service requirements) studies begin in 2025, followed by Stage 2 architectural consolidation through 2026, and Stage 3 protocol freeze targeted for 2027, positioning Release 20 as the dedicated study phase for 6G, with Release 21 marking the start of normative 6G specifications. This progression is aligned with the ITU-R IMT-2030 framework, which targets early 2029 for technology proposals and 2030 for full system definition. Beyond 3GPP and ITU-R, 6G standardization and pre-standardization efforts span multiple organizations \cite{rfc9315,
etsi_gr_eni_055,etsi_gr_eni_056,etsi_zsm_002,etsi_mec_003,
oran_native_ai_architecture,etsi_gr_isc_003_v1_1_1_2026}. The IETF is advancing intent-based networking, AI-assisted network management, and automation architectures; ETSI is evolving frameworks for experiential networked intelligence (ENI), zero-touch service management (ZSM), and multi-access edge computing (MEC); the O-RAN Alliance is integrating AI-native control loops within open RAN architectures; and IEEE initiatives are exploring sub-THz communications, integrated sensing and communication (ISAC) \cite{etsi_gr_isc_003_v1_1_1_2026}, and next-generation PHY innovations \cite{habib2026generative,yu2026reasoning}.

The ITU IMT-2030 framework \cite{itu2026imt2030} identifies the 2024–2027 period as a critical stage for defining technical performance requirements and evaluation criteria for 6G systems, including AI-native operation, integrated sensing and communication, ultra-low latency, high reliability, and massive-scale connectivity \cite{itu_m2160,3gpp_tr_38914}. A recurring theme across 6G roadmaps is the notion of \emph{AI-native} networking, in which artificial intelligence is not merely an auxiliary optimization tool but a foundational architectural element \cite{itu_m2160,oran_native_ai_architecture,etsi_gr_eni_056}. 3GPP TR 22.870 explicitly identifies AI agents as automated intelligent entities capable of intent understanding, contextual reasoning, self-learning, and collaborative decision-making within the 6G system \cite{3gpp_tr_22870}. In parallel, recent IETF drafts on network AI agents and AI agent communication protocols outline use cases in which centralized orchestration agents decompose high-level intents into subtasks executed by specialized service agents, such as connectivity and QoS assurance agents, operating across core and RAN domains \cite{ietf_draft_tong_6g_agents}. These documents further emphasize the need for standardized agent-to-agent communication, task management, identity governance, low-latency interaction, and cross-domain interoperability \cite{ietf_draft_rosenberg_ai_protocols}. In parallel with 6G standardization efforts, the telecommunications industry is advancing toward L4 high-level autonomous networks, characterized by end-to-end closed-loop intelligence combining real-time perception, active analysis, autonomous decision-making, and automated execution \cite{etsi_zsm_002,ietf_draft_zhao_nma,zte2026autonomous}. Recent industrial reports highlight the 2025–2027 period as a critical phase for achieving fully autonomous network operations, supported by the integration of LLMs, multi-agent systems, and digital twin technologies \cite{zte2026autonomous}. 

Despite this architectural momentum, most current AI deployments in wireless systems remain optimization-centric \cite{3gpp_tr_22870,3gpp_tr_23801_01,ietf_draft_tong_6g_agents}. Deep learning models address well-scoped functions such as beam management, traffic forecasting, and slice selection, while reinforcement learning enhances bounded control loops. Even as Stage-2 6G architectural studies advance, intelligence is typically framed within predefined objective functions and reactive feedback cycles \cite{3gpp_tr_23801_01,3gpp_tr_38914}. However, the IETF agent drafts already anticipate multi-agent coordination, task decomposition, capability exposure, and autonomous decision-making beyond simple KPI optimization. As 6G evolves toward extreme heterogeneity—integrating terrestrial and non-terrestrial domains, sensing–communication convergence, digital twins, immersive services, and cross-vertical autonomy—such narrowly scoped optimization models become insufficient. What emerges instead is the need for reasoning-capable, intent-aware, and collaborative agentic intelligence embedded within the network control fabric.

The transition from 6G study items in Release 20 to normative specifications in Release 21 creates a narrow but decisive architectural window \cite{3gpp_rel20_workplan,3gpp_tr_23801_01}. Decisions taken during the current study phase will shape the role of intelligence in 6G systems for the next decade \cite{itu_m2160}. We contend that this transition must not merely standardize AI-assisted optimization, but rather redefine intelligence as a first-class architectural layer within the network. By elevating intelligence from isolated learning modules to distributed reasoning agents operating above deterministic protocol layers, 6G can evolve beyond reactive optimization toward proactive, intent-aware autonomy \cite{xiao2026towards}. Such a shift transforms the network from a performance-tuning system into a self-reasoning infrastructure capable of anticipating, adapting, and governing its own behavior within complex and heterogeneous operational environments \cite{zhang2026multimodal}.

\textcolor{black}{
The concept of Agentic AI-Native 6G is aligned with emerging standardization directions across 3GPP Release 20/21 study items \cite{3gpp_tr_22870,3gpp_tr_23801_01,3gpp_tr_38914}, IETF drafts on AI agents and intent-based networking \cite{ietf_draft_tong_6g_agents,ietf_draft_rosenberg_ai_protocols,rfc9315}, ETSI ENI/ZSM initiatives \cite{etsi_gr_eni_055,etsi_gr_eni_056,etsi_zsm_002}, and the ITU IMT-2030 vision \cite{itu_m2160,itu2026imt2030}. While these efforts do not yet define a unified agentic architecture, they collectively highlight the growing importance of intent-driven operation, AI-native control, distributed intelligence, and agent-based interaction. In this context, agentic intelligence should be viewed as a forward-looking architectural direction that is compatible with ongoing discussions rather than a feature already standardized.
}

To move beyond purely conceptual architectural discussions and assess the practical feasibility of such agentic intelligence, it is essential to evaluate whether LLM-based agents can operate within the stringent constraints of 6G systems. In particular, deploying these agents across the device--edge--core continuum introduces fundamental tradeoffs between reasoning capability, inference latency, and resource efficiency \cite{nguyen2026agentic}. \textcolor{black}{
This paper advances the central thesis that future 6G networks will require a semantic control plane in which LLM-based agents operate as bounded, policy-governed reasoning entities over deterministic 3GPP infrastructure. In this architecture, intelligence is elevated from function-level optimization to system-level reasoning, while execution remains anchored in telecom-grade enforcement mechanisms. This thesis raises three key questions:
}

\begin{itemize}

\item \textcolor{black}{\textbf{Q1:} What architectural framework is required to integrate agentic intelligence into AI-native 6G networks, enabling LLM-based agents to operate as policy-governed reasoning entities over deterministic 3GPP infrastructure?}

\item \textcolor{black}{\textbf{Q2:} Can LLM-based agents operate effectively under realistic device--edge--core constraints, considering latency, throughput, memory, and inference efficiency in 6G environments?}

\item \textcolor{black}{\textbf{Q3:} What deployment principles emerge from the trade-offs between reasoning capability, latency, throughput, and memory, and how should these trade-offs guide the placement and orchestration of agents across the device--edge--core continuum?}

\end{itemize}

To answer these questions, the main contributions of this paper are as follows:

\begin{itemize}
    \item We propose a novel \textit{Agentic AI-Native 6G paradigm} in which LLM-based agents operate as autonomous, policy-governed, and self-reasoning entities within a semantic control plane layered above deterministic 3GPP infrastructure.

    \item We present a comprehensive \textit{four-layer architecture} that integrates deterministic infrastructure, semantic abstraction, hierarchical agentic reasoning, and a distributed multi-agent fabric spanning the device--edge--core continuum.

    \item We introduce a \textit{semantic control framework} enabling intent-aware, context-driven, and trust-aware decision-making through structured representations of intent, policy, knowledge, and uncertainty.

    \item We develop and implement an \textit{agentic evaluation pipeline} combining OpenClaw orchestration with Ollama/llama.cpp-based local inference to support tool-augmented and multi-step reasoning for LLM agents.

    \item We evaluate a diverse set of original and quantized LLMs on the 6G-Bench benchmark, providing a systematic analysis of trade-offs between reasoning performance, latency, throughput, and memory efficiency, and demonstrating the need for heterogeneous deployment across the device--edge--core continuum.

    \item We identify and analyze key \textit{open challenges and research directions} for Agentic AI-Native 6G, including latency--reasoning trade-offs, cross-layer coordination, trust and governance, and scalable multi-agent deployment, providing a roadmap for realizing practical and trustworthy self-reasoning networks.
\end{itemize}

The remainder of this paper is organized as follows. Section \ref{sec:related} reviews related work on Agentic AI-Native Networks. Section \ref{sec:2} discusses the evolution from optimization-centric AI toward agentic intelligence in 6G networks. Section \ref{sec:3} presents the proposed Agentic AI-Native 6G architecture, detailing its layered semantic control framework and distributed multi-agent design. Section \ref{sec:4} describes the experimental methodology and evaluates the performance of quantized LLM agents using the 6G-Bench benchmark under local inference settings. Section \ref{sec:5} outlines key challenges, trade-offs, and research directions for the deployment of agentic intelligence in 6G systems. Finally, Section \ref{sec:6} concludes the paper and highlights future research opportunities.

\section{Related Work}
\label{sec:related}

The integration of artificial intelligence into wireless networks has evolved from function-level optimization toward more autonomous and adaptive paradigms. While early approaches focused on improving performance metrics through machine learning, recent research explores agentic intelligence as a foundation for next-generation 6G systems. In this section, we review the most relevant works and position our contribution within the emerging landscape of agentic AI-native networking.

\subsection{Agentic AI for 6G Network Management and Orchestration}

Several recent works investigate the role of agentic AI in enabling autonomous network management and orchestration. Khowaja \textit{et al.}~\cite{khowaja2025integration} propose a multi-layer agentic AI framework for mission-critical applications, demonstrating improved response time, resource allocation, and operational efficiency. Their work highlights the benefits of contextual reasoning and adaptive decision-making in dynamic environments. Similarly, Antonopoulos \textit{et al.}~\cite{antonopoulos2025agile} introduce the AGILE-6G framework, which organizes distributed agentic subsystems to support end-to-end service orchestration. Their approach emphasizes collaborative decision-making, self-healing capabilities, and context-aware adaptation across network domains. In the context of industrial systems, Li \textit{et al.}~\cite{li2025future} propose an agentic orchestration framework for cyber--physical digital twins, enabling efficient intent-driven pipeline configuration in smart manufacturing environments. These works collectively demonstrate the potential of agent-based systems for improving automation and adaptability in 6G applications, but they remain largely focused on specific use cases or orchestration layers.

\subsection{Agentic AI Architectures and Networking Frameworks}

Beyond application-specific solutions, several studies propose architectural frameworks for agentic networking. Xiao \textit{et al.}~\cite{xiao2025toward} introduce AgentNet, a framework that enables interaction, knowledge sharing, and collaboration among multiple agents based on generative foundation models. Their work envisions a networking ecosystem composed of autonomous agents capable of task-driven cooperation. Feng \textit{et al.}~\cite{feng2025towards} explore the convergence of semantic communication and agentic intelligence in AI-native RAN systems. They provide a taxonomy of semantic abstraction levels and agent coordination strategies, emphasizing the importance of cross-layer reasoning and semantic-aware representations. Zhang \textit{et al.}~\cite{zhang2026toward} further extend this vision by introducing the concept of agentification in edge intelligence, where distributed entities evolve into autonomous agents capable of perception, reasoning, and continuous adaptation. These works highlight the architectural potential of agentic AI but do not fully integrate such capabilities into a unified network control framework.

\subsection{LLM-Based Agents for Networking and Intent-Driven Systems}

The emergence of large language models (LLMs) has enabled new forms of reasoning and decision-making in networking systems. Jiang \textit{et al.}~\cite{jiang2026agentic} propose a hierarchical multi-agent framework for intent-based networking, where LLM agents translate high-level intents into executable network configurations through iterative reasoning processes. Their work demonstrates improved performance over rule-based and direct LLM approaches. Chatzimiltis \textit{et al.}~\cite{chatzimiltis2026agentic} apply LLM-based agents to automate security compliance in RAN systems using retrieval-augmented generation, enabling explainable validation and automated remediation. Similarly, Li \textit{et al.}~\cite{li2026agentic} investigate agentic intelligence at the physical layer, focusing on intent-aware communication and adaptive link configuration under dynamic conditions.

\subsection{Positioning of This Work}

Despite significant progress, existing works exhibit several limitations. First, many approaches are domain-specific, focusing on individual aspects such as orchestration, RAN optimization, or security, without providing a unified architectural perspective. Second, agentic AI is often treated as an application-layer enhancement rather than a fundamental component of the network control plane. Third, current frameworks lack a formal semantic abstraction that integrates intent, policy, trust, and contextual knowledge into a unified representation for reasoning. Finally, the deployment of LLM-based agents is rarely analyzed from a systems perspective, particularly with respect to latency, efficiency, and scalability across the device--edge--core continuum.

In contrast to prior work, this paper proposes a holistic Agentic AI-Native 6G architecture in which reasoning agents operate as first-class entities within a semantic control plane layered above deterministic network infrastructure. By integrating semantic abstraction, hierarchical reasoning, and distributed multi-agent coordination, the proposed framework moves beyond isolated agent-based solutions toward a unified system-level design for self-reasoning networks. In addition, this work provides an extensive empirical evaluation of LLM agents under realistic deployment constraints, offering new insights into the trade-offs between reasoning capability, latency, and resource efficiency.

\textcolor{black}{
As shown in Table~\ref{tab:agentic_conceptual}, the proposed Agentic AI-Native 6G paradigm differs fundamentally from existing AI-assisted, AI-native, and intent-based networking approaches. While prior paradigms primarily rely on optimization or rule-based automation, agentic 6G introduces a unified semantic state and enables multi-step, tool-grounded reasoning over intent, policy, and context. In addition, coordination shifts from centralized or function-level control to distributed multi-agent interaction across the device--edge--core continuum. Importantly, actions are not executed directly but are validated and enforced through policy-governed interfaces, enabling bounded autonomy while preserving deterministic network guarantees.
}

\begin{figure*}[t]
    \centering
    \includegraphics[width=0.82\textwidth]{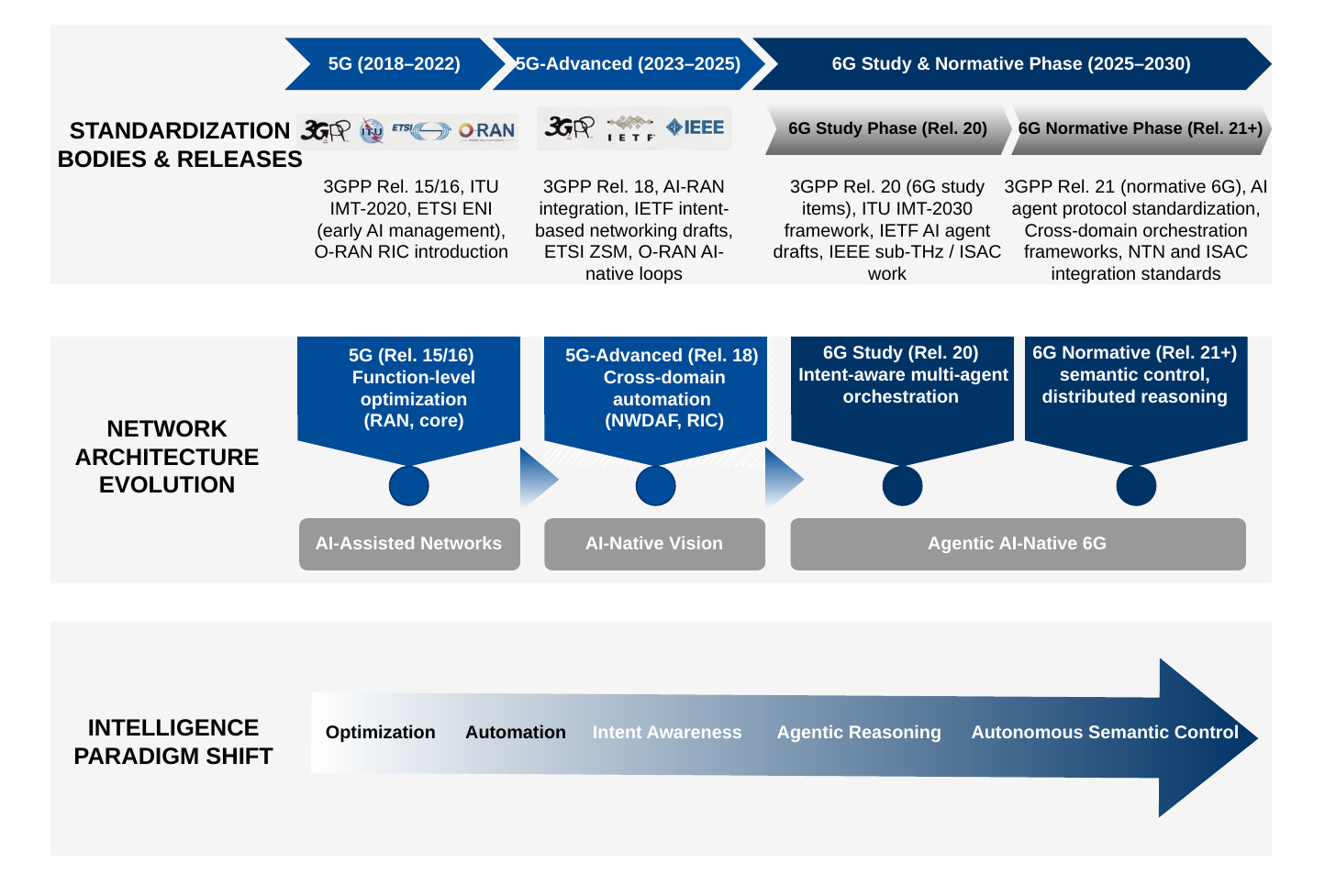}
    \caption{Evolution from 5G optimization-centric AI to Agentic AI-Native 6G aligned with global standardization milestones. The progression from 3GPP Rel-15/16 through Rel-18 and Rel-20/21 reflects the structural shift from KPI-driven optimization to semantic, LLM-augmented multi-agent reasoning embedded within the 6G control plane.}
    \label{fig:evolution_5g_6g}
\end{figure*}

\begin{table*}[t]
\centering
\caption{\textcolor{black}{Comparison of Agentic AI-Native 6G with AI-native, intent-based, and autonomous networking approaches.}}
\label{tab:agentic_conceptual}
\renewcommand{\arraystretch}{1.2}
\begin{tabular}{p{2cm} p{2.5cm} p{2.5cm} p{3cm} p{4.5cm}}
\toprule
\textbf{Dimension} & \textbf{AI-Assisted} & \textbf{AI-Native} & \textbf{Intent/Autonomous} & \textbf{Agentic AI-Native 6G} \\
\midrule

\textcolor{black}{\textbf{Decision Logic}} 
& \textcolor{black}{KPI optimization} 
& \textcolor{black}{Learned optimization} 
& \textcolor{black}{Intent translation} 
& \textcolor{black}{Goal-driven reasoning} \\

\midrule

\textcolor{black}{\textbf{Semantic State}} 
& \textcolor{black}{None} 
& \textcolor{black}{Limited} 
& \textcolor{black}{Intent + context} 
& \textcolor{black}{Intent + policy + trust + uncertainty} \\

\midrule

\textcolor{black}{\textbf{Reasoning}} 
& \textcolor{black}{None} 
& \textcolor{black}{Reactive} 
& \textcolor{black}{Rule-based} 
& \textcolor{black}{LLM-based multi-step reasoning} \\

\midrule

\textcolor{black}{\textbf{Coordination}} 
& \textcolor{black}{Function-level} 
& \textcolor{black}{Cross-domain} 
& \textcolor{black}{Centralized} 
& \textcolor{black}{Distributed multi-agent} \\

\midrule

\textcolor{black}{\textbf{Execution}} 
& \textcolor{black}{Direct control loops} 
& \textcolor{black}{AI-assisted control} 
& \textcolor{black}{Workflow automation} 
& \textcolor{black}{Tool-mediated, policy-governed} \\

\midrule

\textcolor{black}{\textbf{Autonomy}} 
& \textcolor{black}{Low} 
& \textcolor{black}{Medium} 
& \textcolor{black}{High (closed-loop)} 
& \textcolor{black}{Bounded, validated autonomy} \\

\bottomrule
\end{tabular}
\end{table*}

\begin{table*}[t]
\centering
\caption{\textcolor{black}{Evolution from Optimization-Centric AI to Agentic AI-Native 6G (System and Performance Perspective)}}
\label{tab:agentic_evolution_structured}
\renewcommand{\arraystretch}{1.2}
\begin{tabular}{p{3cm} p{6.2cm} p{6.2cm}}
\toprule
\textbf{Dimension} & \textbf{AI-Assisted / AI-Enabled Networks} & \textbf{Agentic AI-Native 6G} \\
\midrule

\textbf{Primary Role of AI} 
& • Function-level optimization \newline
  • Cross-domain automation (SON, NWDAF, RIC) 
& • Intent-aware multi-agent orchestration \newline
  • Semantic reasoning across RAN, core, edge \\

\midrule

\textbf{Decision Logic} 
& • Fixed KPIs and reward functions \newline
  • Bounded control loops 
& • Goal interpretation + task decomposition \newline
  • Policy and trust-aware validation \\

\midrule

\textbf{Latency Orientation} 
& • 10–20 ms eMBB average \newline
  • $\sim$1 ms URLLC air interface 
& • 0.1–1 ms IMT-2030 research targets \newline
  • 25 $\mu$s edge packet processing (dUPF) \\

\midrule

\textbf{Reliability Target} 
& • $\sim 10^{-5}$ BLER (eMBB) \newline
  • $10^{-5}$–$10^{-6}$ URLLC 
& • $10^{-7}$ research reliability target \newline
  • Zero packet loss at 100 Gbps (edge dUPF demo) \\

\midrule

\textbf{User Plane Architecture} 
& • Centralized UPF \newline
  • Partial edge breakout 
& • Distributed UPF (dUPF) at edge \newline
  • AI-DN local breakout for AI traffic \\

\midrule

\textbf{Throughput Capability} 
& • Tens of Gbps (5G peak) \newline
  • Domain-level slicing 
& • 50–200 Gbps IMT-2030 research peak \newline
  • 100 Gbps hardware-accelerated edge validation \\

\midrule

\textbf{Session Scalability} 
& • Standard PDU session management \newline
  • Centralized scaling 
& • 60,000 UE sessions demonstrated \newline
  • 1,000 sessions/sec setup rate \\

\midrule

\textbf{Hardware Acceleration} 
& • CPU-based packet processing \newline
  • Limited offload 
& • BF3 DPU accelerated pipelines (GTP, QER, URR, FAR) \newline
  • 3.2 TB/s Grace memory bandwidth \\

\midrule

\textbf{Energy Efficiency} 
& • General-purpose server processing \newline
  • Higher CPU load under scaling 
& • Full user-plane offload to DPU \newline
  • Operation with only 2 Grace CPU cores \\

\midrule

\textbf{AI Workload Support} 
& • KPI optimization workloads \newline
  • Reactive automation 
& • XR, VSS, V2X, real-time AI conversations \newline
  • Micro-outage and jitter-sensitive sessions \\

\bottomrule
\end{tabular}
\\
\textit{Abbreviations:} 
UPF: User Plane Function; 
dUPF: Distributed User Plane Function; 
AI-DN: AI-specific Data Network; 
SON: Self-Organizing Network; 
NWDAF: Network Data Analytics Function; 
RIC: RAN Intelligent Controller; 
URLLC: Ultra-Reliable Low-Latency Communications; 
BLER: Block Error Rate; 
IMT-2030: International Mobile Telecommunications 2030 framework; 
DPU: Data Processing Unit; 
GTP: GPRS Tunneling Protocol; 
QER: QoS Enforcement Rule; 
URR: Usage Reporting Rule; 
FAR: Forwarding Action Rule; 
VSS: Video Search and Summarization; 
V2X: Vehicle-to-Everything; 
XR: Extended Reality.
\end{table*}

\section{From Optimization to Agentic Intelligence}
\label{sec:2}

The integration of artificial intelligence into wireless networks has evolved through successive layers of automation. Each generation has expanded the scope of decision-making—from parameter tuning at the physical layer to orchestration across distributed network functions \cite{mehmood2026bridging}. However, as 6G enters formal study phases within 3GPP and parallel discussions within IETF emphasize AI agents and intent-driven interaction, a deeper architectural distinction becomes necessary: the difference between optimizing predefined metrics and reasoning over intent, policy, trust, and long-horizon consequences.

Figure~\ref{fig:evolution_5g_6g} illustrates this evolution from 5G optimization-centric AI to Agentic AI-Native 6G, aligned with major standardization milestones across 3GPP, ITU, IETF, ETSI, and O-RAN. The progression from 3GPP Releases~15/16 through~18 and the ongoing Releases~20/21 phases reflects a structural shift—from function-level KPI optimization to cross-domain automation, and ultimately to semantic, LLM-augmented multi-agent reasoning embedded within the 6G control plane. Table~\ref{tab:agentic_evolution_structured} complements this qualitative roadmap with quantitative performance shifts. The transition from AI-assisted optimization to agentic AI-native 6G is not merely conceptual; it reflects measurable changes in latency targets, reliability envelopes, throughput scales, session density, and infrastructural architecture.

\subsection{AI-Assisted Networks: Metric-Driven Optimization}

As depicted in the leftmost stage of Fig.~\ref{fig:evolution_5g_6g}, early AI-assisted 5G deployments focused on function-level optimization within deterministic protocol abstractions \cite{prabhashana2025machine}. Machine learning has enhanced specific network functions, including beam selection, interference coordination, mobility management, traffic prediction, and power allocation \cite{mhatre2025transfer}. In early AI-assisted deployments, machine learning enhanced specific network functions such as beam selection, interference coordination, mobility management, traffic prediction, and power allocation \cite{li2025reinforcement}. These systems operated within predefined optimization objectives—minimizing latency, maximizing throughput, or reducing packet loss—under fixed protocol abstractions \cite{ghafouri2026toward}.

As indicated in Table~\ref{tab:agentic_evolution_structured}, AI-assisted systems were aligned with 5G eMBB performance envelopes, including average latencies on the order of 10–20~ms and block error rates around $10^{-5}$. Intelligence was largely confined to function-level optimization loops and to centralized user-plane architectures \cite{zhang2025joint}. While effective for improving spectral efficiency and throughput, this paradigm did not account for semantic intent, policy constraints, or cross-domain orchestration. AI served as an optimizer rather than a deliberative decision-maker \cite{sanjalawe2025review}.

\subsection{AI-Enabled Networks: Closed-Loop Automation}

With 5G-Advanced (Rel-18), illustrated in the second stage of Fig.~\ref{fig:evolution_5g_6g}, AI expanded into cross-domain automation. AI expanded into cross-domain automation. SON, NWDAF, and RIC-based control loops enabled feedback-driven orchestration across RAN and core domains. This phase supported URLLC targets approaching $\sim$1~ms air-interface latency and reliability levels between $10^{-5}$ and $10^{-6}$, as reflected in Table~\ref{tab:agentic_evolution_structured}.

However, despite tighter latency and reliability constraints, intelligence remained bounded by predefined reward structures and state representations. Automation improved SLA compliance and congestion mitigation, but it remained reactive. Even partial edge breakout did not guarantee deterministic continuity in AI-driven interactive sessions, which are sensitive to jitter and micro-outages. The network optimized performance metrics, but it did not consider goals.

\subsection{Agentic AI-Native Networks: Semantic and Long-Horizon Reasoning}

\textcolor{black}{
Agentic AI-Native 6G is envisioned to introduce both qualitative and quantitative shifts.} IMT-2030 research targets push latency toward 0.1–1~ms and reliability toward $10^{-7}$ levels. Peak throughput expectations extend toward 50–200~Gbps, while connection density scales toward $10^{6}$–$10^{8}$ devices per km$^2$. These targets significantly exceed previous optimization envelopes and impose new architectural requirements.

As shown in Table~\ref{tab:agentic_evolution_structured}, enabling agentic workloads requires distributed user plane architectures. Recent demonstrations of distributed UPF (dUPF) at the network edge achieved a packet-processing latency of 25~$\mu$s and sustained 100~Gbps throughput with zero packet loss \cite{nvidia2025dupfnativeai6g}. Furthermore, 60,000 UE sessions were supported at a setup rate of 1,000 sessions per second, with full hardware acceleration using DPU-based pipelines and only two Grace CPU cores.

These figures illustrate the infrastructural substrate required for large-scale agentic AI services, including XR, video search and summarization (VSS), vehicle-to-everything (V2X), and real-time conversational AI. Such services are highly sensitive to micro-outages, token-stream jitter, and session instability—performance dimensions that are not captured by traditional average-throughput metrics. \textcolor{black}{
In the agentic paradigm, decisions are expected to be evaluated not solely by instantaneous KPI improvements}, but by alignment with higher-level semantic abstractions, including service intent, regulatory policy, trust relationships, and long-horizon risk. An agent operates over a semantic state space integrating telemetry, intent, policy context, trust status, and environmental uncertainty. Unlike automation—which reacts within bounded control loops—agency introduces deliberation. An agent interprets goals, decomposes tasks, evaluates counterfactual outcomes, invokes network APIs, coordinates with peer agents, and validates actions against safety constraints. Intelligence thus becomes architectural rather than functional.

The quantitative progression captured in Table~\ref{tab:agentic_evolution_structured} therefore reflects more than increased performance. It marks a structural transformation of the network—from a reactive optimization engine operating at 10–20~ms latency scales to a distributed, edge-accelerated, semantic control fabric capable of microsecond packet processing and sub-millisecond intent-aware orchestration. \textcolor{black}{
In this vision, 6G is not merely faster connectivity, but is increasingly being discussed as a self-reasoning infrastructure capable of governing its own adaptation under extreme heterogeneity, distributed intelligence, and mission-critical autonomy.
}

\begin{figure*}[t]
    \centering
    \includegraphics[width=0.81\textwidth]{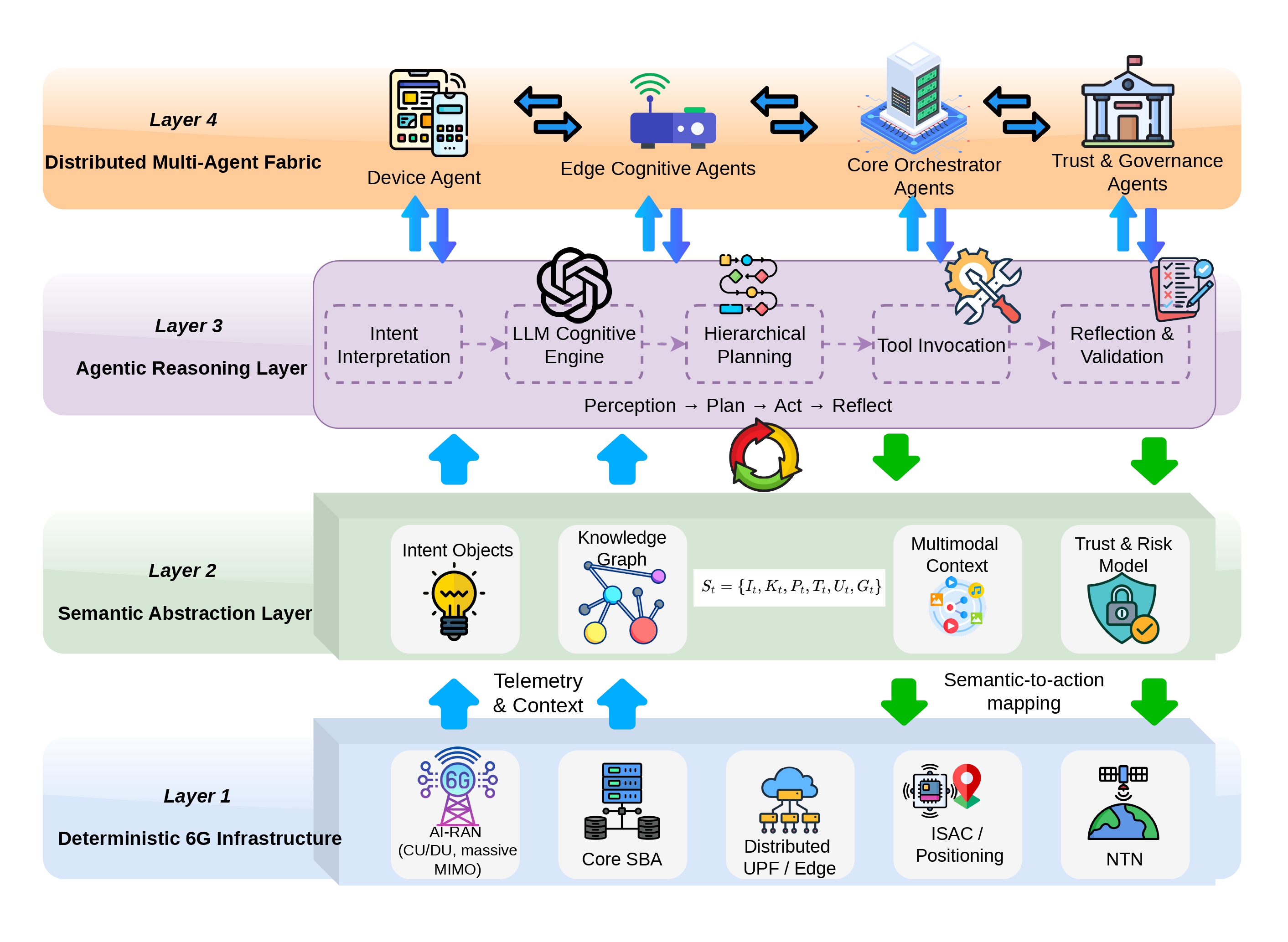}
    \caption{Layered semantic control architecture for Agentic AI-Native 6G. A deterministic 6G infrastructure substrate (Layer~1) provides telecom-grade enforcement, while a semantic abstraction layer (Layer~2) represents intent, cross-domain context, and trust/risk state. An LLM-augmented agentic reasoning layer (Layer~3) performs bounded intent interpretation, task decomposition, planning, tool invocation, and reflection. A distributed multi-agent fabric (Layer~4) coordinates device, edge, core, and governance agents under identity and trust mechanisms, enabling tool-grounded, policy-constrained adaptation without violating deterministic guarantees.}
    \label{fig:agentic6g_architecture}
\end{figure*}

\begin{figure*}[t]
    \centering
    \includegraphics[width=0.9\textwidth]{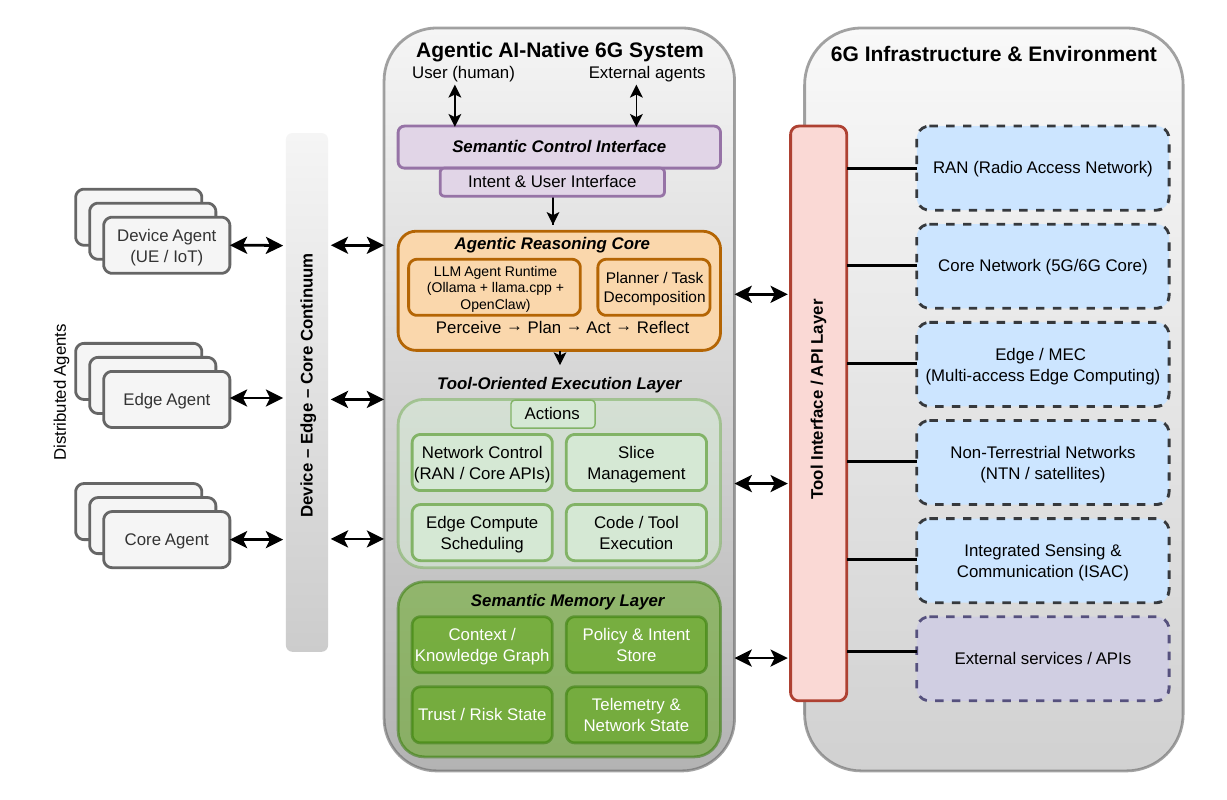}
    \caption{System-level architecture of an Agentic AI-Native 6G system. The figure illustrates semantic control, LLM-based reasoning (Ollama + llama.cpp + OpenClaw), tool-oriented execution, and semantic memory, along with distributed agents operating across the device–edge–core continuum and their interaction with 6G infrastructure components, including RAN, core network, edge/MEC, non-terrestrial networks, and external services.}
    \label{fig:agentic_6g_system}
\end{figure*}

\section{LLM Agents as Reasoning Entities in Agentic 6G}
\label{sec:3}

Building upon the quantitative evolution summarized in Table~\ref{tab:agentic_evolution_structured} and ongoing 3GPP and IETF discussions on AI agents in 6G \cite{draftStephanAI6G,draftYuAI6G}, we propose a layered semantic control architecture for Agentic AI-Native 6G, in which LLM-based agents operate as bounded, policy-governed reasoning entities embedded above deterministic network execution, as presented in Fig. \ref{fig:agentic6g_architecture}.

While Fig. \ref{fig:agentic6g_architecture} presents a layered conceptual architecture of Agentic AI-Native 6G, Fig. \ref{fig:agentic_6g_system} provides a system-level realization of this framework. Specifically, it illustrates the internal structure of an agentic 6G system, including the semantic control interface for intent representation, an LLM agent runtime (\textit{Ollama + llama.cpp + OpenClaw}) enabling multi-step reasoning and tool interaction, and a planning and coordination module for task decomposition and multi-agent orchestration. The architecture further incorporates a tool-oriented execution layer supporting network control, slice management, and edge compute scheduling, as well as a semantic memory layer capturing context, policy, trust, and telemetry information. In addition, the figure highlights the interaction between distributed agents across the device--edge--core continuum and the underlying 6G infrastructure, including RAN, core network, edge/MEC, non-terrestrial systems, and external services, interconnected through standardized control and API interfaces. This system-level view bridges the gap between abstract architectural layers and the practical deployment of agentic intelligence in 6G networks.

\subsection{Layer 1 - Infrastructure Layer}

The infrastructure layer constitutes the formally specified execution fabric of 6G. It preserves deterministic timing, protocol correctness, and interoperability while incorporating advanced enablers, including AI-RAN, disaggregated CU/DU stacks, multi-band radio operation, accelerated user-plane processing, and edge-cloud convergence \cite{kundu2025ai}. Agentic intelligence does not alter the correctness of this substrate; rather, it relies upon its stability, predictability, and enforceable constraints \cite{khan2023ai}.

\subsubsection{Layer 1.1 - Radio Access and AI-RAN Fabric}

The radio access fabric integrates software-defined RAN (vRAN) \cite{kim2025cost,lou2026ad}, AI-accelerated CU/DU architectures \cite{Samsung2026CUDU}, massive MIMO \cite{fu2026local}, and multi-band operation across FR1 and FR2 \cite{ETSI381012}. Disaggregation enables flexible deployment across centralized and edge environments, while AI-assisted PHY/MAC components (e.g., neural channel estimation or adaptive scheduling) enhance spectral efficiency within bounded latency envelopes \cite{Abebe2025AIPhysicalLayer}. Despite increasing programmability, radio operations remain deterministic. Sub-millisecond scheduling, mobility management, and radio resource control continue to operate under strict real-time constraints \cite{kim2025large}.

\subsubsection{Layer 1.2 -  Core Network and Service-Based Architecture}

The core domain operates through a service-based architecture in which modular network functions expose standardized APIs for orchestration, control, and capability discovery \cite{sun2025advancing}. This architecture enables network slicing and differentiated connectivity, policy control and subscriber management, analytics and network exposure services, as well as secure authentication and authorization frameworks that govern access to resources \cite{ojaghi2026intent}. These control primitives provide the formal foundation for translating intent-driven service requests and cross-domain orchestration into enforceable configurations across RAN, edge, and core domains \cite{zhang2025explainable}. Despite increasing programmability and exposure of network capabilities, policy enforcement, session anchoring, and control-plane signaling remain strictly specified, bounded, and auditable, ensuring that architectural flexibility and agent-driven adaptation do not compromise deterministic behavior, interoperability, or regulatory compliance \cite{ali2025ai}.

\subsubsection{Layer 1.3 -  Edge Compute and Distributed User Plane}

Edge computing infrastructures and distributed User Plane Functions (dUPF) form the latency-critical data-plane substrate of Agentic 6G \cite{harkous2026flat}. By enabling local breakout and proximity-based processing for AI-intensive workloads, this domain supports deterministic packet forwarding under strict timing and reliability envelopes. Hardware-accelerated data-plane pipelines provide high-throughput packet handling, low and predictable forwarding latency, session scalability under dense-UE conditions, and a clear separation between control and data paths. These capabilities anchor AI-DN traffic flows, enable split inference placement across device–edge–core hierarchies, and preserve session continuity under mobility and dynamic slice adaptation \cite{vashisht2026multimodal}. As interactive and multimodal AI services become increasingly sensitive to jitter and micro-outages, this edge-accelerated substrate serves as the foundational performance layer upon which higher-level semantic reasoning and agentic coordination depend \cite{zhang2026multimodal,boateng2026move}.

\subsubsection{Layer 1.4 -  Sensing, Positioning, and Non-Terrestrial Extensions}

The 6G infrastructure expands beyond conventional connectivity by embedding environmental perception and extended coverage capabilities directly into the network fabric. Integrated sensing and communication (ISAC) \cite{zhang2026integrated}, high-precision positioning services, reconfigurable intelligent surfaces (RIS) \cite{hua2026implementing}, and non-terrestrial network (NTN) \cite{sacchi2026ecosystemic} extensions collectively enlarge both the perceptual and geographical reach of the system. These technologies generate continuous streams of contextual data and introduce new operational degrees of freedom for mobility management, coverage optimization, and cross-domain coordination. Despite this increased flexibility, deterministic correctness, reliability guarantees, and security enforcement remain invariant across the infrastructure layer. Agentic intelligence may request adaptive reconfiguration of beams, slices, compute placement, or sensing policies, but execution always proceeds through validated, policy-bound, and standardized control mechanisms that preserve telecom-grade stability and compliance \cite{jiang2026large}.

\subsection{Layer 2 -  Semantic Abstraction Layer}

Positioned above the execution of the deterministic protocol, the semantic abstraction layer transforms heterogeneous network signals and service goals into structured machine-interpretable state representations. This layer is the cognitive bridge between infrastructure-level telemetry and agent-level reasoning. Rather than exposing raw metrics alone, it encodes meaning, constraints, and relationships across communication, sensing, compute, and policy domains.

\subsubsection{Layer 2.1 - Intent and Service Semantics}

\textcolor{black}{
6G is expected to introduce intent-based interaction paradigms} in which users, operators, or third-party applications articulate high-level goals without specifying the underlying technical implementation \cite{wang2025survey}. Within this semantic layer, intent is formalized as structured objects that encode service objectives—such as latency bounds, bandwidth targets, and reliability guarantees—alongside resource constraints including energy budgets, coverage domains, and mobility context. Regulatory requirements, compliance rules, and multi-stakeholder policy conditions are embedded directly into these semantic representations, ensuring that objectives are evaluated within enforceable boundaries \cite{alemany2025defining}. By elevating intent to a first-class semantic entity decoupled from low-level configuration parameters, the network can systematically translate natural-language expressions or API-based requests into validated, policy-constrained operational directives that remain compatible with deterministic infrastructure execution \cite{chen20266gagentgym,tong2026wirelessbench}.

\subsubsection{Layer 2.2 - Cross-Domain Context and Knowledge Graph}

6G systems operate across tightly coupled and dynamically interacting domains, including RAN, core, edge compute, integrated sensing and communication (ISAC), non-terrestrial extensions (NTN) \cite{jing2025llm}, and application ecosystems \cite{shi2026stacked}. Within this semantic sublayer, telemetry and operational state from these heterogeneous domains are consolidated into structured knowledge graphs that explicitly encode interdependencies and causal relationships. Rather than treating KPIs as isolated indicators, the network models correlations among radio load and edge compute utilization, mobility dynamics, and slice reconfiguration, as well as sensing inputs and trajectory prediction, and user behavior patterns and differentiated connectivity policies. By embedding these relationships into a unified contextual representation, the system supports holistic, cross-layer reasoning and anticipatory decision-making, thereby enabling coordinated adaptation at the system level instead of fragmented local optimization \cite{zheng2026ai}.

\subsubsection{Layer 2.3 - Multimodal, Sensing, and Digital Twin Integration}

Beyond conventional communication telemetry, 6G significantly broadens the spectrum of semantic inputs available to the network. Integrated sensing and communication (ISAC), high-precision positioning services, RIS-assisted propagation insights, and digital twin environments collectively embed environmental awareness into network cognition \cite{tran2025network}. This semantic sublayer consolidates multimodal inputs—including structured text, telemetry streams, and audio–video signals—with real-time sensing and positioning feeds, predictive digital twin simulations, and externally retrieved knowledge sources, such as retrieval-augmented generation pipelines \cite{vashisht2026multimodal}. By fusing these heterogeneous modalities into a coherent contextual representation, the network can anticipate system evolution, evaluate counterfactual scenarios, and validate candidate actions prior to execution. Such predictive and pre-emptive reasoning becomes essential for mission-critical services where safety, reliability, and temporal precision must be preserved under dynamic environmental conditions.

\subsubsection{Layer 2.4 - Trust, Risk, and Uncertainty Modeling}

Operational decisions in 6G must explicitly account for uncertainty, adversarial risk, and cross-domain trust boundaries \cite{arana2025explainable,mahmood2025securing}. This semantic sublayer embeds agent identity and authentication state, authorization levels, inter-domain trust relationships, confidence scores associated with telemetry and inference outputs, and predicted congestion, anomaly likelihood, and fault probability directly into the system state representation \cite{afaq2026ibn}. By internalizing these trust and risk attributes, reasoning agents can evaluate not only performance trade-offs but also systemic exposure, reliability envelopes, and compliance implications \cite{ferrag20266g}. A unified semantic state at time $t$ can therefore be expressed as

\begin{equation}
S_t = \{I_t, K_t, P_t, T_t, U_t, G_t\},
\end{equation}

where $I_t$ denotes structured intent, $K_t$ captures cross-domain knowledge and telemetry, $P_t$ encodes policy constraints, $T_t$ represents trust and identity state, $U_t$ models uncertainty and risk indicators, and $G_t$ reflects multimodal and digital twin context. This abstraction separates semantic meaning from raw protocol variables, enabling higher-layer reasoning and risk-aware deliberation while preserving deterministic enforcement within the infrastructure layer.

\subsection{Layer 3: Agentic Reasoning Layer}

The agentic reasoning layer hosts LLM-augmented cognitive entities that can interpret intent, decompose complex objectives, coordinate across heterogeneous domains, and invoke network capabilities through standardized interfaces. Unlike optimization-centric control loops bound to predefined reward functions, this layer performs structured deliberation over semantic state, balancing communication, compute, sensing, trust, and energy constraints under bounded latency budgets \cite{wu2025agentic}.

\subsubsection{Layer 3.1 - Intent Interpretation and Goal Decomposition}

Agentic reasoning is initiated by translating structured intent into actionable, domain-specific objectives. High-level goals—such as ultra-reliable XR delivery, mobility-safe V2X coordination, or energy-aware IoT orchestration—are systematically decomposed into coordinated sub-tasks spanning RAN, core, edge compute, sensing, and user-plane domains. This decomposition requires cross-layer trade-off analysis, for example, balancing radio load against edge compute capacity, selecting appropriate slices for differentiated connectivity, determining whether classical or neural PHY/MAC modes should be activated within AI-RAN \cite{khan2023ai}, and incorporating NTN or ISAC resources when environmental awareness or extended coverage is necessary. By transforming abstract intent into structured execution plans while preserving policy constraints and SLA guarantees \cite{yungaicela2026rslaq}, this sublayer enables deliberative orchestration without compromising deterministic enforcement boundaries.

\subsubsection{Layer 3.2 - Hierarchical and Distributed Planning}

To comply with the sub-millisecond latency envelopes and strict reliability guarantees of 6G systems, deliberation within the reasoning layer must be hierarchically distributed. Latency-critical decisions—such as mobility-triggered slice steering, edge inference placement, or rapid beam adjustments—are executed close to the RAN or edge domains, where timing determinism is preserved. In contrast, longer-horizon planning tasks—including capacity forecasting, cross-domain coordination, NTN resource integration, and predictive SLA assurance—are delegated to core-level orchestration agents with broader system visibility. This hierarchical structure enables dynamic placement of inference across device, edge, and cloud tiers; arbitration of GPU and accelerator resources shared between AI-RAN and reasoning workloads \cite{polese2026beyond}; reconfiguration of massive MIMO parameters under mobility \cite{buzzi2026user}; and adaptive RIS-assisted beam management in response to environmental changes. Planning depth, therefore, becomes adaptive rather than static, continuously balancing reasoning complexity against latency budgets, reliability envelopes, and energy constraints.

\subsubsection{Layer 3.3 - Tool-Oriented Execution and Network Control Integration}

Reasoning outcomes are never applied directly to the network; instead, they are translated into structured and constrained tool invocations that target standardized control interfaces. These tools include RAN Intelligent Controllers (xApps and rApps) \cite{li2026multi}, network slicing and orchestration APIs, edge compute schedulers, distributed UPF control primitives, and exposure functions for sensing and positioning services. By grounding decisions in well-defined service-based interfaces, the architecture preserves compatibility with 3GPP service-based principles and existing policy enforcement mechanisms. Actions are parameterized, versioned, and logged, ensuring traceability and auditability while preventing direct manipulation of low-level protocol variables. This tool-oriented abstraction layer thus mediates between cognitive reasoning and deterministic execution, enabling adaptive behavior without compromising control-plane integrity or interoperability.

\begin{figure*}[t]
    \centering
    \includegraphics[width=\textwidth]{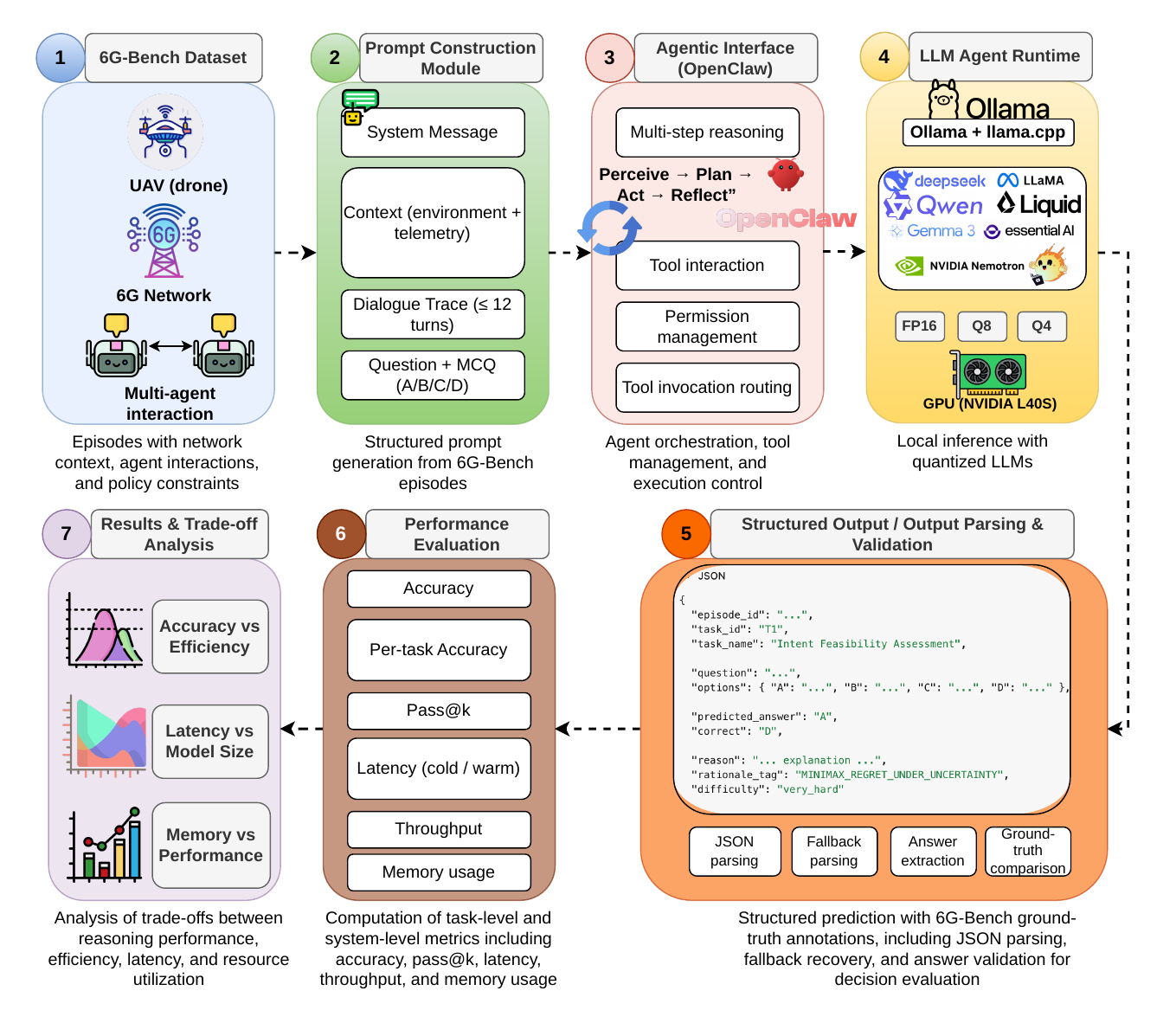}
    \caption{Evaluation pipeline of quantized LLM agents on 6G-Bench under local inference. Structured benchmark episodes are converted into prompts and processed through an OpenClaw-based agentic interface, which orchestrates multi-step reasoning, tool interaction, permission management, and invocation routing. Inference is executed locally through Ollama and llama.cpp-based runtimes across original and quantized model variants, and the resulting outputs are parsed, validated against 6G-Bench ground truth, and analyzed using both task-level and system-level metrics, including accuracy, pass@k, latency, throughput, and memory usage.}
    \label{fig:evaluation_pipeline}
\end{figure*}

\subsubsection{Layer 3.4 - Reflection, Validation, and Safety Guardrails}

Following execution, outcomes are continuously evaluated against predicted effects, reliability envelopes, and policy constraints to ensure that adaptive behavior remains aligned with deterministic guarantees \cite{deng2026ai}. This reflective stage incorporates digital twin simulation \cite{tran2025network} for counterfactual verification, real-time KPI monitoring and anomaly detection, cross-agent consensus mechanisms in collaborative scenarios, and formal validation against SLA commitments and regulatory boundaries \cite{tang2026guardrailing}. Guardrail mechanisms enforce strict separation between reasoning generation and execution authorization, ensuring that decisions violating latency budgets, compliance requirements, or risk thresholds are rejected, revised, or safely rolled back \cite{shahid2026iot}. Operationally, each agent adheres to a bounded cognitive cycle comprising semantic perception from Layer~2, contextual memory updates, hierarchical planning and task decomposition, tool-grounded execution via Layer~1 interfaces, and reflective validation with adaptive refinement. Through this structured feedback loop, 6G evolves from a reactive optimization framework into a deliberative, policy-aware, and risk-sensitive control fabric capable of reasoning over long-horizon stability, cross-layer dependencies, and mission-level objectives without compromising telecom-grade reliability \cite{alves20256g}.

\subsection{Layer 4: Distributed Multi-Agent Fabric}

The deployment of agents in 6G is intrinsically distributed across heterogeneous domains, including devices, RAN/edge infrastructures \cite{mahmoud2026review}, core networks, and external application ecosystems. Unlike centralized AI control architectures, Agentic 6G relies on coordinated multi-agent collaboration, underpinned by standardized mechanisms for discovery, identity, and communication.

\subsubsection{Layer 4.1 - On-Device and Embedded Agents}

On-device agents are embedded in user equipment, vehicles, wearables, robots, XR terminals, and IoT nodes, serving as the primary interface for multimodal perception, intent articulation, and localized inference \cite{shahid2026iot}. Operating under stringent energy, thermal, and computational constraints, these agents initiate intent-driven service invocation, perform local pre-processing and privacy-preserving inference, and participate in collaborative decision-making with network-based agents \cite{bi2025opportunities}. They also enable horizontal inter-device coordination in multi-agent scenarios, forming dynamic communication domains across heterogeneous access technologies. However, due to resource limitations, complex reasoning and large-model execution are typically offloaded via split inference mechanisms to edge or core domains \cite{xiao2026understanding}. Consequently, robust identity management, secure authentication, and cross-domain trust enforcement are foundational requirements for ensuring safe and reliable device–network collaboration within the distributed agent fabric.

\subsubsection{Layer 4.2 - Edge Cognitive Agents}

Edge cognitive agents are deployed in close proximity to AI-RAN infrastructures, MEC platforms, and distributed UPF instances, positioning them at the convergence point of radio control, compute resources, and user-plane forwarding \cite{zhao2025world}. This locality enables deterministic low-latency adaptation and continuity for time-sensitive services. Edge agents manage AI-DN local breakout and traffic steering, perform mobility-aware slice adaptation, coordinate split inference execution across device and cloud domains, and arbitrate shared GPU or accelerator resources between AI-RAN workloads and higher-layer reasoning tasks. They also mitigate micro-outages and jitter for interactive, multimodal services whose performance is highly sensitive to transient disruptions \cite{wang2025cognitive}. Operating under bounded reasoning depth, edge agents prioritize latency-critical decisions while remaining aligned with global intent, policy constraints, and orchestration directives established by higher-layer agents, thereby preserving both responsiveness and systemic coherence.

\subsubsection{Layer 4.3 - Core Orchestrator and Cross-Domain Agents}

Core orchestrator agents operate with system-wide visibility across communication, sensing, compute, and non-terrestrial domains, enabling coordinated decision-making beyond the scope of localized edge control. Positioned at the intersection of RAN, core, MEC, ISAC, and NTN infrastructures, these agents perform long-horizon planning and translate high-level intent into coherent cross-domain service compositions \cite{wei2025large}. Their responsibilities encompass cross-slice and cross-tenant orchestration, coordination between terrestrial and non-terrestrial resources, integrated sensing and communication allocation, predictive SLA assurance, capacity forecasting, and controlled exposure of network knowledge to support retrieval-augmented AI workflows \cite{feng2025towards}. By reconciling competing objectives—such as latency versus energy efficiency, or throughput versus fairness—under multi-stakeholder policy constraints, core agents ensure that distributed adaptations remain globally consistent and strategically aligned with operator and regulatory requirements.

\subsubsection{Layer 4.4: Trust, Governance, and Resilience Agents}

A dedicated supervisory stratum underpins trust and systemic stability within the distributed multi-agent fabric. Because reasoning agents may influence slice configurations, beamforming policies, compute placement decisions, or user-plane routing behaviors, execution must remain subject to strict authorization, validation, and policy enforcement. Governance agents, therefore, implement cross-domain identity verification and credential management, enforce role-based authorization and explicit consent mechanisms, coordinate fault detection and redundancy strategies, and perform load balancing to prevent systemic bottlenecks \cite{ferrag2026alphasec}. For high-impact or safety-critical actions, digital twin–assisted pre-validation provides predictive assurance before deterministic enforcement, while anomaly-detection pipelines monitor for hallucination-induced inconsistencies or adversarial perturbations. By maintaining a structural separation between cognitive inference and execution authorization, this layer preserves telecom-grade reliability guarantees while still enabling adaptive, autonomous behavior across heterogeneous 6G environments \cite{bang2025hallulens}.

\begin{table}[t]
\centering
\caption{Experimental setup and evaluation configuration.}
\label{tab:experimental_setup}
\renewcommand{\arraystretch}{1.1}
\scriptsize
\begin{tabular}{p{1.2cm} p{2.4cm} p{3.8cm}}
\toprule
\textbf{Category} & \textbf{Setting} & \textbf{Value} \\
\midrule

\multirow{5}{*}{\textbf{Dataset}}
& Source scenarios & 113{,}475 (from $\alpha^3$-Bench) \cite{ferrag2026alpha}\\
& Generated pool & 10{,}000 MCQs \cite{ferrag20266g} \\
& Evaluation set & 3{,}722 MCQs (validated) \\
& Tasks & 30 (T1--T30) across 5 groups \\
& Format & Very-hard MCQ (4 choices, 1 correct, multi-step reasoning) \\

\midrule
\multirow{3}{*}{\textbf{Input}}
& Dialogue length & $\leq 12$ AI-agent interaction turns \\
& Network features & Latency (ms), Jitter (ms), Throughput (Mb/s), Packet loss (\%), Edge load \\
& 6G semantic features & Network slice (URLLC/eMBB/mMTC), Intent, Policy constraints, Trust level, SLA requirements \\

\midrule
\multirow{3}{*}{\textbf{Models}}
& Count & 20 models \\
& Scale & 135M--27B parameters \\
& Context & 8K--262K tokens \\

\midrule
\multirow{2}{*}{\textbf{Quantization}}
& Formats & Q4, Q4\_K\_M, Q8 \\
& Reduction & $\sim$2--4$\times$ memory \\

\midrule
\multirow{3}{*}{\textbf{Runtime}}
& Framework & Ollama (v0.17.7) + llama.cpp \\
& Orchestration & OpenClaw (agentic orchestration layer) \\
& Mode & Local (no API) \\

\midrule
\textbf{Hardware}
& GPU & NVIDIA L40S (46 GB VRAM) \\
& CUDA & Version 13.0 \\

\midrule
\multirow{2}{*}{\textbf{Decoding}}
& Accuracy & Temperature $T = 0$ \\
& Pass@k & Temperature $T = 0.7$, $k \in \{3,5\}$ \\

\midrule
\multirow{4}{*}{\textbf{Metrics}}
& Accuracy & [0,1] \\
& Latency & ms (cold, warm) \\
& Throughput & tokens/s \\
& Memory & GB (peak VRAM) \\

\midrule
\multirow{2}{*}{\textbf{Evaluation}}
& Runs & Single inference per query; diagnostic profiling on one query repeated 20 times ($\mu,\sigma$) \\
& Scope & Full evaluation set (3{,}722 validated MCQs) \\
\bottomrule
\end{tabular}
\end{table}

\section{Evaluation of Quantized LLM Agents}
\label{sec:4}

\subsection{Experimental Setup}

To evaluate the feasibility of deploying LLM-based agents within Agentic AI-Native 6G, we conduct an extensive experimental study using the 6G-Bench benchmark \cite{ferrag20266g} under local inference conditions. The evaluation jointly considers reasoning capability and systems-level efficiency, enabling a comprehensive analysis of the tradeoffs introduced by model quantization. Figure~\ref{fig:evaluation_pipeline} summarizes the end-to-end evaluation workflow used in this study. Starting from structured 6G-Bench episodes, each benchmark instance is transformed into a unified prompt containing the system message, contextual information, truncated dialogue trace, and multiple-choice question. The prompt is then processed through an OpenClaw-based agentic interface that manages multi-step reasoning and tool-oriented interaction, while local inference is executed through Ollama and llama.cpp-based runtimes on the evaluated LLM models. The resulting predictions are parsed and validated against the benchmark ground truth, after which both task-level and system-level metrics are computed to quantify the trade-offs between reasoning quality, latency, throughput, and memory efficiency across different model and quantization configurations.

\subsubsection{Benchmark and Task Formulation}

We utilize 6G-Bench \cite{ferrag20266g}, a domain-specific benchmark designed to assess reasoning and decision-making capabilities in network-centric and mission-driven scenarios. The benchmark comprises structured episodes that represent UAV-assisted 6G environments, each capturing dynamic interactions among agents, network conditions, and policy constraints.

Each episode is transformed into a compact textual representation through a structured summarization pipeline. This includes (i) initial system context, comprising environment conditions, airspace constraints, UAV state, and policy information, and (ii) a truncated dialogue trace containing agent interactions, tool invocations, observations, and network indicators such as latency, jitter, packet loss, throughput, and edge load. In our experiments, the dialogue trace is truncated to at most 12 turns in order to bound prompt length while preserving the most relevant contextual information.

The evaluation task is formulated as a multiple-choice question (MCQ) problem. For each episode, models are provided with a task definition, the summarized episode context, and a question with four candidate answers (A/B/C/D). The benchmark covers 30 task categories spanning intent feasibility, conflict resolution, slice selection and switching, trust-aware offloading, SLA prediction, multi-agent coordination, interoperability, security response, and other 6G agentic decision-making functions.

\subsubsection{Prompting and Output Parsing}

All models are evaluated using a unified prompting template consisting of a system message and a user message. The system message defines the model as an expert 6G network AI agent evaluator, while the user message includes the target task identifier and definition, the summarized episode context, the question text, and the four answer options. To ensure consistent evaluation, models are instructed to return only a valid JSON object of the form $\texttt{"answer"}: "A/B/C/D"$. Predictions are extracted automatically through structured JSON parsing. When strict JSON parsing fails, a fallback parser is applied to recover the answer label from raw model output. The final parsed answer is then compared against the benchmark ground-truth label.

\begin{table*}[t]
\centering
\caption{Evaluated LLM Models, Quantization, and Alignment with Agentic AI-Native 6G Deployment}
\scriptsize
\label{tab:evaluated_models}
\renewcommand{\arraystretch}{1.15}
\begin{tabular}{l l l l l l l l l}
\toprule
\textbf{Model} & \textbf{Family} & \textbf{Params} & \textbf{Variant} & \textbf{Quant.} & \textbf{Arch. Type} & \textbf{Ctx} & \textbf{Alignment} & \textbf{Target} \\
\midrule

deepseek-r1-1.5b & DeepSeek-R1 & 1.5B & Distilled & -- 
& Distilled Dense & 131K & Lightweight Reasoning Agent & Device \\

deepseek-r1-1.5b-q4\_K\_M & DeepSeek-R1 & 1.5B & Quantized & Q4\_K\_M 
& Distilled Dense & 131K & Ultra-light Agent & Device \\

gemma3:4b & Gemma 3 & 4B & Original & -- 
& Dense (Multimodal) & 128K & General Agent & Edge \\

gemma3:4b-it-q4\_K\_M & Gemma 3 & 4B & Instruction-tuned & Q4\_K\_M 
& Dense (Multimodal) & 128K & Efficient Agent & Edge \\

gemma3:4b-it-q8 & Gemma 3 & 4B & Instruction-tuned & Q8 
& Dense (Multimodal) & 128K & High-Quality Agent & Edge \\

granite4:350m & Granite 4 & 350M & Original & -- 
& Tool-tuned Dense & 131K & Embedded Agent & Device \\

granite4:350m-h-q8 & Granite 4 & 350M & Quantized & Q8 
& Tool-tuned Dense & 131K & Ultra-light Agent & Device \\

granite4:1b & Granite 4 & 1B & Original & -- 
& Tool-tuned Dense & 131K & Embedded Agent & Device \\

granite4:1b-h-q8 & Granite 4 & 1B & Quantized & Q8 
& Tool-tuned Dense & 131K & Efficient Embedded Agent & Device \\

lfm2:24b & LFM 2 & 24B & Original & -- 
& MoE (2B active) & 32K & Global Reasoning Agent & Core \\

lfm2:24b-q4\_K\_M & LFM 2 & 24B & Quantized & Q4\_K\_M 
& MoE (2B active) & 32K & Efficient Core Agent & Core \\

lfm2:24b-q8 & LFM 2 & 24B & Quantized & Q8 
& MoE (2B active) & 32K & High-Performance Agent & Core \\

llama3.2:1b & LLaMA 3.2 & 1B & Original & -- 
& Dense & 60K & General Agent & Edge \\

llama3.2:1b-instruct-q4 & LLaMA 3.2 & 1B & Instruction-tuned & Q4 
& Dense & 60K & Lightweight Agent & Device \\

llama3.2:3b & LLaMA 3.2 & 3B & Original & -- 
& Dense & 60K & General Agent & Edge \\

nemotron-3-nano:4b & Nemotron 3 & 4B & Original & -- 
& MoE (A3B) & 262K & High-Efficiency Agent & Edge \\

qwen3.6:27b & Qwen 3.6 & 27B & Original & -- 
& Hybrid (DeltaNet + MoE) & 262K & High-End Multimodal Agent & Core \\

rnj-1:8b & RNJ-1 & 8B & Instruction-tuned & -- 
& Dense & 32K & Reasoning Agent & Edge \\

smollm2:135m & SmolLM 2 & 135M & Original & -- 
& Dense (Compact) & 8K & Ultra-light Agent & Device \\

smollm2:135m-instruct-q4 & SmolLM 2 & 135M & Instruction-tuned & Q4 
& Dense (Compact) & 8K & Ultra-light Agent & Device \\

smollm2:135m-instruct-q8 & SmolLM 2 & 135M & Instruction-tuned & Q8 
& Dense (Compact) & 8K & Efficient Ultra-light Agent & Device \\

\bottomrule
\end{tabular}

\vspace{2mm}
\textit{Abbreviations:} 
Params: number of parameters; 
Quant.: quantization scheme; 
Arch. Type: model architecture type; 
Ctx: context length (maximum input tokens); 
MoE: Mixture-of-Experts; 
A3B: active-parameter routing configuration; 
Q4, Q4\_K\_M, Q8: quantization formats used in Ollama/llama.cpp-based runtimes; 
Target: intended deployment domain (Device, Edge, Core).

\end{table*}

\subsubsection{Evaluated Models}

We evaluate a diverse set of open-weight LLMs spanning multiple model families, architectural paradigms, and parameter scales, as summarized in Table~\ref{tab:evaluated_models}. The evaluated models include DeepSeek-R1 \cite{deepseekai2025deepseekr1incentivizingreasoningcapability}, Gemma~3 \cite{kamath2025gemma3}, Granite~4 \cite{mishra2024granite}, LFM~2 \cite{liquidai2025lfm2}, LLaMA~3.2 \cite{grattafiori2024llama}, Nemotron \cite{parmar2024nemotron}, Qwen~3.6 \cite{qwen3.6-27b}, RNJ \cite{rnj1_instruct}, and SmolLM~2 \cite{allal2025smollm2smolgoesbig}, covering dense, distilled, tool-tuned, mixture-of-experts (MoE), and hybrid architectures. For each model family, we consider both the original and quantized variants, including GGUF-based configurations such as Q4, Q4\_K\_M, and Q8, thereby enabling a systematic analysis of compression–performance trade-offs. The selected models span a wide range of parameter sizes, from compact models with fewer than $200$M parameters (e.g., SmolLM~2) to large-scale models with more than $20$B parameters (e.g., Qwen~3.6 and LFM~2), thereby providing broad coverage of deployment scenarios. In addition to parameter scale, the evaluated models differ significantly in context length, architectural design, and training objectives. Notably, long-context and hybrid architectures, such as Qwen~3.6 and Nemotron, enable extended reasoning over large context windows, whereas compact models, such as SmolLM~2, are optimized for efficient on-device execution. These characteristics allow us to analyze the suitability of different model classes for deployment across the device–edge–core continuum in Agentic AI-Native 6G systems.

\subsubsection{Inference Environment}

All experiments are conducted using the Ollama framework (version 0.17.7) in a fully local deployment setting (\texttt{http://localhost:11434}), without reliance on external APIs. The evaluation leverages llama.cpp-based runtimes for executing quantized models (e.g., GGUF formats such as Q4, Q4\_K\_M, and Q8), enabling efficient local inference across a range of model sizes. In addition, the inference pipeline is interfaced through OpenClaw, which provides an agentic orchestration layer supporting multi-step reasoning, tool interaction, and structured agent workflows aligned with the proposed multi-agent architecture. The underlying hardware platform consists of a multi-GPU system equipped with NVIDIA L40S GPUs, each providing 46~GB of VRAM, running CUDA~13.0 with NVIDIA driver version 580.105.08. This configuration enables controlled evaluation of both full models and quantized variants under consistent runtime conditions while preserving reproducibility and precise measurement of latency, throughput, and memory usage.

\subsubsection{Performance Metrics}

We collect both task-level and systems-level metrics for each evaluated model. Task-level metrics include overall accuracy, per-task accuracy, and pass@k for selected reasoning-intensive tasks. Systems-level metrics include total inference latency, model load duration, token generation duration, output throughput, and peak GPU memory usage. To better reflect deployment conditions, we distinguish between cold-start latency, which includes model-loading overhead, and warm-inference latency, which excludes loading and reflects steady-state execution after the model is resident in memory. Pass@k is evaluated only on a selected subset of tasks involving ambiguity, conflict resolution, trust-aware decision-making, interoperability, and security-sensitive reasoning, where multiple stochastic attempts provide a meaningful measure of robustness.

Let $\mathcal{Q}$ denote the set of all evaluated benchmark questions, with $|\mathcal{Q}| = N$. For each question $q \in \mathcal{Q}$, let $\hat{y}_q$ be the predicted answer and $y_q$ the ground-truth answer. The indicator function $\mathbb{I}(\cdot)$ returns 1 if its argument is true and 0 otherwise.

\begin{itemize}

\item \textbf{Overall Accuracy:}
\begin{equation}
\mathrm{Acc} = \frac{1}{N} \sum_{q \in \mathcal{Q}} \mathbb{I}(\hat{y}_q = y_q),
\end{equation}
where $\mathcal{Q}$ is the set of all benchmark questions with cardinality $N = |\mathcal{Q}|$, $q$ denotes an individual question, $\hat{y}_q$ is the predicted answer, $y_q$ is the ground-truth answer, and $\mathbb{I}(\cdot)$ is the indicator function, which equals 1 if its argument is true and 0 otherwise.

\item \textbf{Per-Task Accuracy:}
\begin{equation}
\mathrm{Acc}_t = \frac{1}{N_t} \sum_{q \in \mathcal{Q}_t} \mathbb{I}(\hat{y}_q = y_q),
\end{equation}
where $\mathcal{Q}_t \subseteq \mathcal{Q}$ is the subset of questions associated with task $t$, $N_t = |\mathcal{Q}_t|$ is the number of questions in that task, and the remaining symbols follow the definitions given in the overall accuracy metric.

\item \textbf{Pass@k:}
\begin{equation}
\mathrm{Pass@}k(q) = \mathbb{I}\!\left(\exists i \in \{1,\dots,k\} \text{ such that } \hat{y}_q^{(i)} = y_q \right),
\end{equation}
\begin{equation}
\mathrm{Pass@}k = \frac{1}{|\mathcal{Q}^{(k)}|} \sum_{q \in \mathcal{Q}^{(k)}} \mathrm{Pass@}k(q),
\end{equation}
where $\hat{y}_q^{(i)}$ denotes the prediction obtained at the $i$-th stochastic attempt for question $q$, $k$ is the number of attempts, and $\mathcal{Q}^{(k)} \subseteq \mathcal{Q}$ is the subset of questions eligible for multi-attempt evaluation. The remaining symbols follow the definitions introduced in the overall accuracy metric.

\item \textbf{Latency:}
\begin{equation}
L_{\mathrm{cold}} = T_{\mathrm{total}}, \quad
L_{\mathrm{warm}} = T_{\mathrm{total}} - T_{\mathrm{load}},
\end{equation}
where $T_{\mathrm{total}}$ is the total inference time, $T_{\mathrm{load}}$ is the model loading time, $L_{\mathrm{cold}}$ denotes the cold-start latency including model initialization overhead, and $L_{\mathrm{warm}}$ denotes the steady-state inference latency after the model is loaded.

\item \textbf{Throughput:}
\begin{equation}
\mathrm{Thr} = \frac{n_{\mathrm{out}}}{T_{\mathrm{eval}}},
\end{equation}
where $n_{\mathrm{out}}$ is the number of generated output tokens and $T_{\mathrm{eval}}$ is the token evaluation duration (in seconds), yielding throughput in tokens per second.

\item \textbf{Peak Memory Usage:}
\begin{equation}
M_{\mathrm{peak}} = \max_{\tau \in \mathcal{T}} M(\tau),
\end{equation}
where $M(\tau)$ denotes the GPU memory usage at sampling time $\tau$, $\mathcal{T}$ is the set of sampling instants during the inference execution, and $M_{\mathrm{peak}}$ represents the maximum observed memory usage over the entire inference period.

\item \textbf{Statistical Aggregation:}
\begin{equation}
\mu_x = \frac{1}{R} \sum_{r=1}^{R} x_r,
\end{equation}
\begin{equation}
\sigma_x = \sqrt{\frac{1}{R-1} \sum_{r=1}^{R} (x_r - \mu_x)^2},
\end{equation}
where $\{x_1, x_2, \dots, x_R\}$ denotes the set of repeated measurements of a given metric, $R$ is the number of repetitions, $\mu_x$ is the sample mean, and $\sigma_x$ is the sample standard deviation. This formulation is used for repeated profiling experiments (e.g., latency and throughput), whereas accuracy metrics are computed deterministically.

\end{itemize}

\subsubsection{Execution Protocol}

Each model is evaluated independently over the full 6G-Bench dataset \cite{ferrag20266g} using a unified prompting and decoding pipeline. For standard accuracy evaluation, inference is performed with deterministic decoding ($T=0$) and fixed seeds derived from the episode and task identifiers to ensure reproducibility. For pass@k evaluation, repeated stochastic attempts are generated with a temperature of 0.7 and distinct seeds for each attempt. For every query, detailed runtime metadata are collected, including total latency, load duration, prompt evaluation statistics, token generation statistics, output throughput, and peak VRAM usage. This unified protocol enables a fair comparison between the original and quantized models while preserving reproducibility and consistency in both quality and systems-level measurements.

\begin{table}[t]
\centering
\caption{Reasoning Performance of Evaluated LLM Models on 6G-Bench}
\label{tab:accuracy_results}
\scriptsize
\renewcommand{\arraystretch}{1}
\begin{tabular}{l c c c}
\toprule
\textbf{Model} & \textbf{Accuracy} & \textbf{Pass@3} & \textbf{Pass@5} \\
\midrule

\multicolumn{4}{l}{\textbf{DeepSeek}} \\
DeepSeek-R1-1.5B & 0.422 & 0.693 & 0.807 \\
DeepSeek-R1-1.5B-Q4\_K\_M & 0.422 & 0.693 & 0.807 \\

\midrule
\multicolumn{4}{l}{\textbf{Gemma}} \\
Gemma3-4B & 0.641 & 0.627 & 0.626 \\
Gemma3-4B-Q4\_K\_M & 0.641 & 0.627 & 0.626 \\
Gemma3-4B-Q8 & 0.654 & 0.655 & 0.654 \\

\midrule
\multicolumn{4}{l}{\textbf{Granite}} \\
Granite4-1B & 0.460 & 0.618 & 0.683 \\
Granite4-1B-Q8 & 0.544 & 0.672 & 0.771 \\
Granite4-350M & 0.346 & 0.526 & 0.632 \\
Granite4-350M-Q8 & 0.318 & 0.556 & 0.696 \\

\midrule
\multicolumn{4}{l}{\textbf{LFM}} \\
LFM2-24B & 0.765 & 0.790 & 0.799 \\
LFM2-24B-Q4\_K\_M & 0.765 & 0.790 & 0.799 \\
LFM2-24B-Q8 & \textbf{0.774} & 0.800 & 0.807 \\

\midrule
\multicolumn{4}{l}{\textbf{LLaMA}} \\
LLaMA3.2-1B & 0.358 & 0.489 & 0.598 \\
LLaMA3.2-1B-Q4 & 0.242 & 0.257 & 0.272 \\
LLaMA3.2-3B & 0.627 & 0.740 & 0.790 \\

\midrule
\multicolumn{4}{l}{\textbf{Nemotron}} \\
Nemotron-3-Nano-4B & 0.662 & \textbf{0.839} & \textbf{0.883} \\

\midrule
\multicolumn{4}{l}{\textbf{Qwen}} \\
Qwen3.6-27B & 0.771 & 0.830 & 0.850 \\

\midrule
\multicolumn{4}{l}{\textbf{RNJ}} \\
RNJ-1-8B & 0.700 & 0.826 & 0.875 \\

\midrule
\multicolumn{4}{l}{\textbf{SmolLM}} \\
SmolLM2-135M & 0.224 & 0.237 & 0.255 \\
SmolLM2-135M-Q4 & 0.224 & 0.409 & 0.464 \\
SmolLM2-135M-Q8 & 0.224 & 0.230 & 0.254 \\

\bottomrule
\end{tabular}

\vspace{2mm}
\textit{Metrics:} Accuracy is computed under deterministic decoding (temperature $=0$). Pass@k is evaluated using stochastic decoding on selected reasoning-intensive tasks.
\end{table}

\subsection{Benchmark Performance}
Table~\ref{tab:accuracy_results} summarizes the reasoning performance of all evaluated models on 6G-Bench \cite{ferrag20266g} under both deterministic and multi-attempt inference. A clear hierarchy emerges across model families and scales. Large and optimized architectures achieve the highest overall performance, with RNJ-1-8B reaching the best deterministic accuracy of $0.700$ and a strong pass@5 of $0.875$, while Nemotron-3-Nano-4B exhibits the highest multi-attempt performance with pass@5 of $0.883$ despite slightly lower accuracy ($0.662$). Similarly, LFM2-24B models demonstrate strong and stable performance, achieving up to $0.774$ accuracy and $0.807$ pass@5, indicating that large-scale and mixture-of-experts architectures provide both strong single-shot reasoning and robust recovery through iterative inference. In contrast, smaller models such as LLaMA3.2-1B ($0.358$ accuracy) and SmolLM2-135M ($0.224$ accuracy) show limited reasoning capability, although they benefit from multi-attempt inference to varying degrees (e.g., LLaMA3.2-1B improves to $0.598$ pass@5). These results highlight a clear tradeoff between model scale and reasoning capability, with larger or more advanced architectures achieving superior performance, particularly in multi-step decision-making scenarios.

The impact of quantization, however, is highly model-dependent and reveals several non-trivial behaviors. In some cases, quantization preserves or even improves performance. For instance, Granite4-1B-Q8 improves accuracy from $0.460$ to $0.544$ and pass@5 from $0.683$ to $0.771$, while LFM2-24B-Q8 slightly improves accuracy from $0.765$ to $0.774$. Conversely, other models exhibit negligible or no changes, such as DeepSeek-R1-1.5B, where quantization has no impact on performance ($0.422$ accuracy and $0.807$ pass@5 in both cases), and Gemma3-4B, which remains largely unchanged across all quantization levels (accuracy $\approx 0.64$ and pass@5 $\approx 0.63$–$0.65$). In contrast, aggressive quantization can significantly degrade performance, as observed in LLaMA3.2-1B-Q4, where accuracy drops from $0.358$ to $0.242$ and pass@5 from $0.598$ to $0.272$. Interestingly, for very small models such as SmolLM2-135M, quantization has mixed effects: while accuracy remains unchanged ($0.224$), the Q4 variant significantly improves pass@5 from $0.255$ to $0.464$, suggesting enhanced stochastic exploration despite limited base capability. These findings demonstrate that quantization is not a uniform compression strategy but rather a system-level tradeoff that can preserve, degrade, or even enhance reasoning behavior depending on the model architecture and inference dynamics.

\begin{table*}[t]
\centering
\caption{Inference Profiling of Evaluated LLM Models (Single-Query Analysis)}
\scriptsize
\label{tab:profiling_models_grouped}
\renewcommand{\arraystretch}{1}
\begin{tabular}{l c c c c c}
\toprule
\textbf{Model} & \textbf{Cold latency (ms)} & \textbf{Warm latency (ms)} & \textbf{Throughput (tok/s)} & \textbf{VRAM (GB)} & \textbf{Input / Output} \\
\midrule

\multicolumn{6}{l}{\textbf{DeepSeek }} \\
DeepSeek-R1-1.5B & 6302 & 1626 & 448.1 & 5.36 & 2427 / 7 \\
DeepSeek-R1-1.5B-Q4\_K\_M & 6374 & 1607 & 454.4 & 5.36 & 2427 / 7 \\

\midrule
\multicolumn{6}{l}{\textbf{Gemma }} \\
Gemma3-4B & 564 & 380 & 183.7 & 6.86 & 2033 / 7 \\
Gemma3-4B-Q4\_K\_M & 8640 & 385 & 190.5 & 6.86 & 2033 / 7 \\
Gemma3-4B-Q8 & 9168 & 404 & 133.8 & 8.39 & 2033 / 7 \\

\midrule
\multicolumn{6}{l}{\textbf{Granite }} \\
Granite4-1B & 5412 & 190 & 181.9 & 14.02 & 1823 / 7 \\
Granite4-1B-Q8 & 4957 & 242 & 187.2 & 4.83 & 1823 / 7 \\
Granite4-350M & 4848 & \underline{136} & 491.1 & 2.31 & 1823 / 7 \\
Granite4-350M-Q8 & 4863 & 171 & 316.8 & 2.64 & 1823 / 7 \\

\midrule
\multicolumn{6}{l}{\textbf{LFM }} \\
LFM2-24B & 9501 & 411 & 139.0 & 14.74 & 2058 / 8 \\
LFM2-24B-Q4\_K\_M & \underline{381} & 348 & 188.8 & 14.75 & 2058 / 8 \\
LFM2-24B-Q8 & 9905 & 426 & 125.8 & 24.94 & 2058 / 8 \\

\midrule
\multicolumn{6}{l}{\textbf{LLaMA }} \\
LLaMA3.2-1B & 237 & \underline{151} & 433.2 & 6.16 & 1835 / 7 \\
LLaMA3.2-1B-Q4 & 4921 & 152 & \underline{590.9} & 5.65 & 1835 / 7 \\
LLaMA3.2-3B & 324 & 236 & 249.9 & 16.83 & 1835 / 7 \\

\midrule
\multicolumn{6}{l}{\textbf{Nemotron }} \\
Nemotron-3-Nano-4B & 5907 & 5813 & 198.0 & 9.82 & 2829 / 10 \\

\midrule
\multicolumn{6}{l}{\textbf{Qwen }} \\
Qwen3.6-27B & 32647 & 32516 & 33.7 & 40.47 & 1931 / 1000 \\

\midrule
\multicolumn{6}{l}{\textbf{RNJ }} \\
RNJ-1-8B & 7472 & 499 & 115.5 & 9.54 & 2014 / 8 \\

\midrule
\multicolumn{6}{l}{\textbf{SmolLM }} \\
SmolLM2-135M & 147 & 123 & 742.6 & 1.07 & 2102 / 11 \\
SmolLM2-135M-Q4 & \textbf{142} & \textbf{117} & \textbf{764.7} & \textbf{0.89} & 2102 / 11 \\
SmolLM2-135M-Q8 & 150 & 123 & 721.9 & 0.94 & 2102 / 11 \\

\bottomrule
\end{tabular}

\vspace{2mm}
\textit{Abbreviations:} 
Cold Lat.: total inference latency including model loading; 
Warm Lat.: steady-state latency excluding load time; 
VRAM: peak GPU memory usage. \textbf{Bold}: best value in column; \underline{Underline}: second-best or notable performance.
\end{table*}

\subsection{Inference Performance and Quantization Analysis}

Quantization plays a central role in enabling the deployment of LLM-based agents under the stringent resource constraints of 6G environments. Table~\ref{tab:profiling_models_grouped} provides a detailed comparison between original and quantized models across latency, throughput, and memory usage, revealing that the benefits of quantization are highly dependent on both model architecture and runtime behavior.

In several cases, quantization significantly improves memory efficiency and computational performance. For example, Granite4-1B-Q8 reduces peak VRAM usage from $14.02$~GB to $4.83$~GB while slightly increasing throughput from $181.9$ to $187.2$~tokens/s. Similarly, SmolLM2-135M-Q4 reduces memory usage from $1.07$~GB to $0.89$~GB and increases throughput from $742.6$ to $764.7$~tokens/s, demonstrating that compact models can benefit substantially from low-bit quantization. These improvements are particularly relevant for device-level and edge deployments, where memory and latency constraints are critical.

However, quantization does not consistently yield improvements across all models. For instance, DeepSeek-R1-1.5B-Q4\_K\_M shows no reduction in memory usage ($5.36$~GB for both original and quantized versions), with only marginal throughput improvement ($448.1$ to $454.4$~tokens/s). Furthermore, quantization may introduce latency variability due to runtime effects, such as model caching and memory allocation strategies. This is evident in the case of LFM2-24B-Q4\_K\_M, which exhibits a substantially lower cold-start latency ($381$~ms) than its original counterpart ($9501$~ms), a behavior attributed to caching rather than to intrinsic model efficiency.

These results indicate that quantization should be viewed as a system-level optimization rather than a uniform compression technique. Its effectiveness depends not only on model size and architecture, but also on runtime characteristics and deployment conditions. In the context of Agentic AI-Native 6G, this suggests that lightweight quantized models are well-suited for device and edge agents, while larger models may still require careful optimization to balance reasoning capability with system efficiency.

\subsection{\textcolor{black}{From Results to Deployment Principles}}

\textcolor{black}{
The experimental results provide concrete insights into how LLM-based agents should be deployed across the device--edge--core continuum in Agentic AI-Native 6G. While the evaluation highlights clear trade-offs between reasoning performance, latency, throughput, and memory usage, these trade-offs also reveal fundamental design principles that directly support the proposed multi-agent architecture.}

\subsubsection{\textcolor{black}{Device-Level Agents (Compact Models):}}
\textcolor{black}{Lightweight models, such as SmolLM2-135M and Granite4-350M variants, achieve the highest efficiency, with low memory usage (below 1~GB VRAM) and high throughput (above 700 tokens/s). However, their reasoning capability remains limited (accuracy around 0.22). These characteristics make them suitable for device-side deployment, where they can perform perception, filtering, lightweight intent classification, and privacy-preserving preprocessing under strict energy and latency constraints.}

\subsubsection{\textcolor{black}{Edge-Level Agents (Mid-Scale Models):}}
\textcolor{black}{Mid-size models, such as RNJ-1-8B, Gemma3-4B, and Nemotron-3-Nano-4B, provide a balanced trade-off between reasoning capability and system efficiency. For example, RNJ-1-8B achieves strong reasoning performance (accuracy 0.700, pass@5 0.875) while maintaining moderate latency and memory requirements. These models are well suited for edge-level deployment, where they can support fast operational decisions, slice adaptation, mobility-aware control, and coordination of device-level agents.}

\subsubsection{\textcolor{black}{Core-Level Agents (Large Models):}}
\textcolor{black}{Large-scale models, such as LFM2-24B and Qwen3.6-27B, achieve the highest reasoning capability but incur significant computational cost, including high latency (hundreds of milliseconds to seconds) and large memory footprints (up to 40~GB VRAM). These models are therefore better suited for core-level deployment, where they can perform long-horizon planning, cross-domain orchestration, policy arbitration, and complex reasoning tasks that are less sensitive to strict latency constraints.}

\subsubsection{\textcolor{black}{Model-Dependent Impact of Quantization:}}
\textcolor{black}{The results demonstrate that quantization does not uniformly preserve performance across models. While some models benefit from quantization (e.g., Granite4-1B-Q8 and LFM2-24B-Q8), others experience negligible gains or even degradation (e.g., LLaMA3.2-1B-Q4). This indicates that quantization must be selected and optimized on a per-model basis, rather than assumed as a universally beneficial compression strategy.}

\subsubsection{\textcolor{black}{Joint Optimization of Agent Placement:}}
\textcolor{black}{Overall, the results show that no single model can simultaneously satisfy all constraints related to reasoning quality, latency, throughput, and memory efficiency. This reinforces the need for a heterogeneous deployment strategy in which agents are distributed across the device--edge--core continuum and selected based on joint optimization of system-level metrics. Such a placement strategy directly supports the proposed layered architecture, where different classes of agents operate at different levels of the network to balance performance and efficiency.}

\section{Key Challenges and Research Directions}
\label{sec:5}

Agentic AI-Native 6G represents a structural transformation across all four architectural layers introduced earlier: deterministic infrastructure (Layer~1), semantic abstraction (Layer~2), agentic reasoning (Layer~3), and distributed multi-agent coordination (Layer~4). Embedding deliberative intelligence into a mission-critical, globally standardized telecom substrate introduces cross-layer tensions that do not arise in conventional AI systems. Unlike cloud-centric AI deployments, 6G systems operate under ultra-low latency constraints, strict reliability envelopes, heterogeneous domain coupling (RAN--core--edge--sensing--NTN), and regulatory accountability. Our empirical results with quantized LLM agents further show that these tensions are not only conceptual, but also measurable in terms of reasoning accuracy, inference latency, throughput, and GPU memory footprint. For example, while large models such as LFM2-24B-Q8 achieve the highest deterministic accuracy of $0.774$, compact models such as SmolLM2-135M-Q4 attain the best efficiency profile, with only $0.89$~GB VRAM usage, $117$~ms warm latency, and $764.7$~tokens/s throughput. Similarly, RNJ-1-8B provides a strong balance between reasoning capability and deployability, reaching $0.700$ accuracy and $0.875$ pass@5, whereas Nemotron-3-Nano-4B achieves the best multi-attempt robustness with pass@5 of $0.883$. These results confirm that realizing a semantic control plane requires addressing architectural challenges that span and interconnect all layers while accounting for the heterogeneous capabilities of different LLM agent classes across the device--edge--core continuum. 

These challenges emerge at a time when early 6G standardization efforts are actively defining system requirements, use cases, and architectural directions through 3GPP Release~20 studies (e.g., TR~22.870, TR~23.801, TR~38.914) \cite{3gpp_tr_22870,3gpp_tr_23801_01,3gpp_tr_38914}, as well as broader international frameworks such as ITU-R M.2160 \cite{itu_m2160}. While these efforts establish the foundation for 6G systems, they do not yet fully capture the implications of embedding agentic, LLM-driven reasoning into the control plane, particularly under strict latency, reliability, and regulatory constraints.

\subsection{Latency--Reasoning Tradeoff}

A fundamental architectural tension arises between the deterministic guarantees of Layer~1 and the deliberative capabilities of Layer~3. The infrastructure layer enforces sub-millisecond scheduling, accelerated user-plane processing, and tightly bounded AI-RAN radio control loops, ensuring predictable timing and reliability envelopes. By contrast, the reasoning layer \textcolor{black}{is expected to introduce} multi-step inference, contextual memory management, and cross-domain planning whose computational complexity scales with problem scope. Our evaluation makes this gap explicit. Even relatively compact reasoning agents exhibit inference latencies that are orders of magnitude above strict radio control timescales: RNJ-1-8B requires $499$~ms warm latency, LFM2-24B-Q8 requires $426$~ms, and Nemotron-3-Nano-4B reaches $5813$~ms under the measured single-query setup. At the opposite end, highly efficient compact models such as SmolLM2-135M-Q4 achieve warm latency of $117$~ms, and LLaMA3.2-1B-Q4 achieves $152$~ms, but their reasoning capabilities remain limited, with deterministic accuracies of $0.224$ and $0.242$, respectively. This mismatch becomes particularly acute in URLLC scenarios — such as V2X coordination, tactile industrial control, or real-time robotics — where decisions on slice steering, beam adaptation, split inference placement, or RIS \cite{hussain2026intent} reconfiguration must meet microsecond-to-millisecond control deadlines. 

The results, therefore, suggest that no single LLM agent class can simultaneously satisfy both high-quality reasoning and extreme timing constraints. High-performing models such as LFM2-24B-Q8 and RNJ-1-8B are better suited for supervisory or orchestration-level reasoning, whereas compact agents such as SmolLM2-135M-Q4, Granite4-350M, or LLaMA3.2-1B-Q4 are more suitable for bounded local support functions at the device or edge. Addressing this tradeoff requires hierarchical reasoning decomposition, where latency-critical inference remains edge-local and bounded while deeper deliberation occurs asynchronously in core domains; speculative execution mechanisms with deterministic fallback policies; pre-validated action templates supported by digital twin simulation in Layer~2.3; and explicit encoding of latency-aware reasoning budgets within the semantic state representation. The central research challenge is to augment control with intelligence without compromising the deterministic safety guarantees that underpin telecom-grade reliability.

This tension is only partially reflected in current standardization efforts, which emphasize ultra-low latency and reliability requirements (e.g., URLLC evolution in TR~38.914 and service requirements in TR~22.870) \cite{3gpp_tr_38914,3gpp_tr_22870}, but do not explicitly account for the computational latency introduced by multi-step AI reasoning. Bridging this gap requires extending existing performance models to incorporate reasoning-aware latency budgets.

\subsection{Cross-Domain Orchestration}

Cross-domain orchestration exposes a fundamental tension between the semantic abstraction layer (Layer~2) and the distributed multi-agent fabric (Layer~4). While Layer~2 consolidates telemetry, ISAC sensing inputs, NTN coverage states, slicing configurations, and compute load into unified semantic representations, Layer~4 disperses agents across devices, edge nodes, core infrastructures, and application domains. Maintaining semantic consistency across these distributed contexts is non-trivial: agents may operate with partial visibility, stale telemetry, heterogeneous policy interpretations, or mobility-induced context shifts. Our evaluation also highlights that cross-domain orchestration cannot assume homogeneous reasoning agents. The measured model space spans ultra-light device-oriented agents such as SmolLM2-135M and Granite4-350M, edge-oriented agents such as Gemma3-4B, Nemotron-3-Nano-4B, and RNJ-1-8B, and core-oriented models such as LFM2-24B and Qwen3.6-27B. These models differ not only in accuracy, but also in context length, architectural style, and runtime profile. For instance, Nemotron-3-Nano-4B provides strong pass@5 robustness ($0.883$) and a large $262$K context window, while Qwen3.6-27B supports similarly long context but incurs $40.47$~GB VRAM and only $33.7$~tokens/s throughput, making it impractical for latency-sensitive multi-agent interaction.

These differences imply that future Agentic 6G systems must support heterogeneous multi-agent fabrics in which device, edge, and core agents possess distinct reasoning depth, memory horizon, and inference budgets. Research challenges, therefore, include reliable cross-domain knowledge synchronization under high mobility and NTN extension, consistent intent interpretation across operator and administrative boundaries, coordinated decision-making in dynamically formed multi-agent groups, and semantic interoperability for multimodal and cross-platform agents. Without standardized mechanisms for discovery, registration, identity binding, policy alignment, and capability advertisement, distributed reasoning risks devolving into fragmented local optimizations rather than coherent system-level orchestration.

This challenge aligns with ongoing work on service-based architectures and multi-domain coordination in 3GPP (TR~23.801) \cite{3gpp_tr_23801_01}, as well as ETSI initiatives such as Zero-touch Service Management (ZSM) \cite{etsi_zsm_002} and Experiential Networked Intelligence (ENI) \cite{etsi_gr_eni_055,etsi_gr_eni_056}, which aim to enable autonomous network operation. However, these frameworks do not yet address heterogeneous LLM-based agent reasoning and dynamic semantic consistency across distributed domains.

\subsection{Energy--Compute Placement Tradeoff}

LLM-augmented reasoning \textcolor{black}{is expected to introduce} significant computational overhead, while 6G roadmaps place strong emphasis on energy efficiency, carbon awareness, and sustainable network operation. This creates a fundamental tradeoff between cognitive sophistication and resource expenditure. Our profiling results show that this tradeoff is strongly model-dependent. For example, Granite4-1B-Q8 reduces VRAM consumption from $14.02$~GB to $4.83$~GB while slightly improving throughput from $181.9$ to $187.2$~tokens/s, making it a more viable candidate for embedded or near-edge deployment. SmolLM2-135M-Q4 similarly reduces memory from $1.07$~GB to $0.89$~GB and improves throughput from $742.6$ to $764.7$~tokens/s, illustrating the efficiency benefits of low-bit quantization for lightweight agents. By contrast, some larger models remain expensive even after quantization: LFM2-24B-Q8 still requires $24.94$~GB VRAM, and Qwen3.6-27B requires $40.47$~GB VRAM with extremely low throughput relative to smaller models.

These observations imply that inference placement across device, edge MEC nodes, and centralized core infrastructures must explicitly account for both reasoning quality and computational footprint. High-capability core agents such as LFM2-24B-Q8 or Qwen3.6-27B may enable deeper planning and long-horizon coordination, but their memory and throughput requirements limit their deployment to well-provisioned environments. Edge-tier models such as RNJ-1-8B, Gemma3-4B-Q8, or Nemotron-3-Nano-4B may offer a more balanced tradeoff, while compact quantized models such as SmolLM2-135M-Q4 and Granite4-1B-Q8 are more plausible for on-device or embedded assistance. Decisions regarding model compression, quantization, mixture-of-experts routing, and adaptive inference scaling must therefore be made in real time to balance performance and energy cost. Minimizing energy per decision cannot come at the expense of SLA compliance, latency guarantees, or reliability envelopes. Consequently, energy-aware meta-reasoning becomes a first-class capability: agents must evaluate not only network state but also their own computational footprint and carbon impact. The tight coupling between the split inference mechanisms in Layer~1.3 and the hierarchical planning strategies in Layer~3.2, therefore, defines a new research frontier in sustainable, compute-network co-designed 6G systems.

Energy efficiency and sustainability are already recognized as key 6G design objectives in ITU-R M.2160 \cite{itu_m2160}, yet current frameworks do not explicitly incorporate the computational cost of AI-driven reasoning. Extending these models to include inference-aware energy metrics is therefore essential.

\begin{table*}[t]
\centering
\caption{\textcolor{black}{Risk categories, representative 6G scenarios, mitigation mechanisms, and security references for agentic AI systems}}
\label{tab:agentic_risks}
\renewcommand{\arraystretch}{1.2}
\scriptsize
\begin{tabular}{p{2.5cm} p{3.5cm} p{4cm} p{3cm}}
\toprule
\textbf{Risk Category} & \textbf{6G Scenario} & \textbf{Mitigation Mechanism} & \textbf{Security Reference} \\
\midrule

\textcolor{black}{Hallucinated decisions} 
& \textcolor{black}{Incorrect slice allocation or QoS policy due to faulty reasoning} 
& \textcolor{black}{Policy validation, guardrails, digital twin pre-validation} 
& \textcolor{black}{CWE-670 (Always-Incorrect Control Flow)} \\

\midrule

\textcolor{black}{Prompt injection / malicious intent} 
& \textcolor{black}{Adversarial input manipulating intent interpretation} 
& \textcolor{black}{Input sanitization, trust-aware filtering} 
& \textcolor{black}{CWE-20 (Improper Input Validation)} \\

\midrule

\textcolor{black}{Agent mis-coordination} 
& \textcolor{black}{Conflicting decisions across edge and core agents} 
& \textcolor{black}{Consensus protocols, hierarchical orchestration} 
& \textcolor{black}{CWE-362 (Race Condition)} \\

\midrule

\textcolor{black}{Latency violation} 
& \textcolor{black}{Delayed decisions affecting URLLC or V2X services} 
& \textcolor{black}{Bounded reasoning, fallback policies} 
& \textcolor{black}{CWE-400 (Uncontrolled Resource Consumption)} \\

\midrule

\textcolor{black}{Unauthorized actions} 
& \textcolor{black}{Agent triggers reconfiguration without authorization} 
& \textcolor{black}{RBAC, identity verification, policy enforcement} 
& \textcolor{black}{CWE-284 (Improper Access Control)} \\

\midrule

\textcolor{black}{Model inconsistency} 
& \textcolor{black}{Unstable behavior due to quantization effects} 
& \textcolor{black}{Model validation and runtime monitoring} 
& \textcolor{black}{CWE-682 (Incorrect Calculation)} \\

\bottomrule
\end{tabular}
\end{table*}

\subsection{Trust, Safety, and Hallucination Risk}

LLM-augmented agents introduce uncertainty modes that are fundamentally absent in deterministic telecom control systems. Hallucinated outputs, inconsistent tool invocation, misinterpreted intent, or context drift may inadvertently alter slice configurations, radio parameters, edge routing policies, or compute placement decisions, potentially propagating instability into Layer~1 execution domains. Our results show that stronger reasoning models are not necessarily the safest in operational settings, because high capability must still be reconciled with deterministic enforcement. For example, RNJ-1-8B, Nemotron-3-Nano-4B, and LFM2-24B variants show the strongest reasoning performance, with accuracy and pass@5 values ranging from $0.700$ to $0.883$, but these are precisely the kinds of agents that could trigger complex, high-impact actions if not tightly constrained. Conversely, low-capability compact agents may have lower action sophistication, but can still introduce errors through inaccurate local decisions, as reflected by the $0.224$ accuracy of SmolLM2-135M variants and the $0.242$ accuracy of LLaMA3.2-1B-Q4.

\textcolor{black}{
Table~\ref{tab:agentic_risks} illustrates how the introduction of LLM-based agents in 6G networks expands the risk landscape beyond traditional protocol-level concerns to include reasoning errors, coordination failures, and adversarial manipulation. These risks highlight the necessity of structured governance mechanisms that enforce policy-constrained execution, validate agent decisions before deployment, and ensure secure and trustworthy interaction across distributed agents. } To mitigate these risks, Layer~4.4 (Trust, Governance, and Verification) must impose structured guardrails over Layer~3 reasoning. This includes a strict separation between reasoning generation and execution authorization, enforcement of policy-constrained action templates, digital twin pre-validation for high-impact operations, cross-agent consensus mechanisms, anomaly-detection pipelines, and role-based access control with explicit consent management. The model-dependent effects of quantization reinforce this need. In some cases, quantization preserves or improves reasoning quality, as with Granite4-1B-Q8 and LFM2-24B-Q8; in others, it substantially degrades it, as with LLaMA3.2-1B-Q4. This means that safety validation cannot assume that a compressed model is merely a cheaper approximation of its original counterpart. The core research challenge lies in designing governance mechanisms that are sufficiently robust to prevent unsafe or non-compliant autonomy while remaining flexible enough to preserve adaptive, context-aware reasoning across heterogeneous 6G environments.

These concerns extend beyond traditional telecom reliability models and are not fully addressed in current specifications, which largely assume deterministic control logic. Emerging ETSI ENI studies on AI-driven networks \cite{etsi_gr_eni_055,etsi_gr_eni_056} begin to explore these issues, but comprehensive governance frameworks for LLM-based agents remain an open challenge.

\subsection{Security and Adversarial Robustness}

Agentic AI-Native 6G significantly broadens the attack surface beyond traditional signaling and protocol-layer vulnerabilities, enabling adversaries to target semantic representations, reasoning processes, and inter-agent coordination mechanisms. Potential threat vectors include prompt injection and semantic manipulation within the abstraction layer (Layer~2), telemetry poisoning and sensing spoofing across ISAC and NTN feeds, malicious agent impersonation within distributed identity frameworks (Layer~4), cross-domain policy exploitation, and compromise of execution tool chains and orchestration interfaces (Layer~3.3). The heterogeneous model landscape identified in our evaluation compounds this challenge. Agent classes differ in architecture, quantization, prompt behavior, and context length, which may produce unequal security exposure. Long-context agents such as Nemotron-3-Nano-4B and Qwen3.6-27B may enable richer contextual reasoning but may also be more vulnerable to context poisoning or adversarial prompt persistence over large windows. Small quantized agents may be easier to deploy widely, but their compressed representations may exhibit unstable or less predictable behavior under adversarial perturbation.

Because reasoning agents can influence slicing decisions, compute placement, and user-plane routing, adversarial perturbations at higher layers may propagate into deterministic control domains. Ensuring robustness, therefore, requires end-to-end security spanning strong agent identity verification, cryptographically protected inter-agent communication, anomaly-aware validation of reasoning outputs, secure discovery and registration mechanisms, and zero-trust execution environments that isolate and constrain tool invocation. A particularly important research direction is the development of security evaluation suites tailored to telecom-grade LLM agents that can test prompt-injection resilience, semantic consistency under partial-telemetry corruption, and adversarial robustness across quantized and full-precision variants. Only through cross-layer security integration can agentic autonomy remain resilient under adversarial conditions.

While security frameworks in telecom systems are well established, they primarily address protocol- and infrastructure-level threats. The introduction of AI agents \textcolor{black}{is expected} to introduce new semantic and reasoning-layer attack surfaces that are not yet covered in existing standards or emerging studies, such as IETF drafts on 6G AI agents \cite{ietf_draft_tong_6g_agents}.

\subsection{Explainability and Regulatory Compliance}

Telecommunications infrastructure operates under stringent regulatory oversight, in which decisions affecting resource allocation, traffic prioritization, lawful interception, or emergency service handling must be auditable, transparent, and legally defensible. In an Agentic AI-Native 6G architecture, this requirement extends directly to the reasoning layer. Layer~3 must therefore generate structured, traceable reasoning artifacts that document the interpretation of intent, task decomposition, and tool invocation paths, while Layer~4.4 enforces logging, certification, authorization, and compliance validation mechanisms before and after execution. Our empirical results underscore that explainability cannot be treated in isolation from deployment heterogeneity. Different model families produce different tradeoffs between capability and operational cost. For instance, RNJ-1-8B and LFM2-24B-Q8 offer strong reasoning quality, but they also operate as higher-capability agents whose outputs may influence broader orchestration decisions. Compact models such as SmolLM2-135M-Q4 or Granite4-1B-Q8 are easier to deploy pervasively, but their limited reasoning fidelity makes post-hoc traceability and bounded-action validation even more important.

The explainability requirement is closely aligned with the intent-based networking principles defined in IETF RFC~9315 \cite{rfc9315}, in which high-level intents must be translated into verifiable network actions. However, the introduction of LLM-based reasoning complicates this mapping, requiring new mechanisms for traceability, validation, and regulatory compliance.

\subsection{Research Directions Across Standardization Phases}

The transition toward Agentic AI-Native 6G must be aligned with ongoing standardization efforts across multiple bodies. 3GPP Release~20 study items (TR~22.870, TR~23.801, TR~38.914) \cite{3gpp_tr_22870,3gpp_tr_23801_01,3gpp_tr_38914,3gpp_rel20_workplan} define the initial 6G service requirements, architectural directions, and system scenarios. Complementary efforts from ITU-R \cite{itu_m2160} establish global objectives, while ETSI initiatives such as ZSM \cite{etsi_zsm_002}, MEC \cite{etsi_mec_003}, ENI \cite{etsi_gr_eni_055,etsi_gr_eni_056}, and ISAC studies \cite{etsi_gr_isc_003_v1_1_1_2026} provide frameworks for automation, edge computing, AI integration, and sensing-aware networks. In parallel, O-RAN Alliance work on native AI architectures \cite{oran_native_ai_architecture} and IETF efforts on intent-based networking and AI agents \cite{rfc9315,ietf_draft_tong_6g_agents} further contribute to the emerging ecosystem.

Addressing the cross-layer challenges of Agentic AI-Native 6G requires a phased research agenda aligned with ongoing 6G study and standardization timelines. In the near-term study phase, foundational work is needed to quantify and bound the interaction between semantic reasoning and deterministic execution. Our results suggest several immediate priorities. First, future benchmarks should move beyond static reasoning quality alone and explicitly evaluate intent-to-action latency under constrained reasoning budgets, since current warm latencies span from $117$~ms for SmolLM2-135M-Q4 to $5813$~ms for Nemotron-3-Nano-4B and more than $32$~s for Qwen3.6-27B in the reported setup. Second, semantic interoperability models should account for heterogeneous agent classes, since high-performing edge and core agents such as RNJ-1-8B, Nemotron-3-Nano-4B, and LFM2-24B variants cannot be treated interchangeably with compact device agents such as SmolLM2-135M or Granite4-350M. Third, energy-aware inference placement strategies should be formalized using measurable resource indicators such as VRAM footprint, throughput, and cold-start overhead. Fourth, digital-twin validation pipelines are needed for safety-critical actions, particularly when quantized models exhibit non-monotonic behavior relative to their original counterparts.

As standardization progresses into the normative phase, research must mature toward formalized architectural mechanisms: standardized agent discovery and identity frameworks, multi-agent communication protocols aligned with 3GPP service-based architecture principles, capability-aware orchestration across AI-RAN, slicing, ISAC, and NTN resources, and governance models that incorporate carbon-aware and sustainability-aware reasoning policies. The evaluation results suggest that this phase should also include standardized agent profiling and certification, so that models can be classified by deployment suitability — e.g., ultra-light device agents, efficient edge agents, and high-capability core agents — based on jointly measured reasoning and system metrics rather than on parameter count alone.

In the longer-term deployment phase, attention shifts toward operational resilience and ecosystem-wide coordination. Self-healing, cross-domain multi-agent systems capable of anticipating cascading failures must be realized, alongside mechanisms for autonomous SLA negotiation and dynamic enforcement under multi-stakeholder policy constraints. Human--AI collaborative semantic control planes will be essential to maintain operator oversight and regulatory compliance while preserving adaptive autonomy. Ultimately, globally interoperable agent communication standards, distributed trust infrastructures, and model lifecycle governance will be required to ensure secure collaboration across operators, application domains, and geopolitical boundaries. The transition to Agentic AI-Native 6G is therefore not simply a matter of scaling AI models; it demands coordinated evolution across deterministic infrastructure, semantic modeling, reasoning algorithms, model compression strategies, and distributed trust mechanisms. Only by resolving these cross-layer tensions can 6G mature into a policy-aware, intent-driven, and safety-preserving intelligent network.

\section{Conclusion}
\label{sec:6}

The transition from 5G-Advanced to 6G entails a fundamental redesign of how intelligence is embedded in network architectures, shifting from optimization-centric AI to agentic, intent-driven systems. In this work, we introduced a four-layer Agentic AI-Native 6G architecture that integrates deterministic infrastructure, semantic abstraction, hierarchical LLM-based reasoning, and distributed multi-agent coordination, enabling policy-aware, cross-domain orchestration while preserving telecom-grade reliability. Beyond this architectural vision, we conducted a comprehensive empirical evaluation of quantized LLM agents using the 6G-Bench benchmark, revealing measurable tradeoffs between reasoning capability, latency, throughput, and memory footprint across device–edge–core deployments. Our results show that high-capability models such as LFM2-24B-Q8 and RNJ-1-8B achieve strong reasoning performance (up to $0.774$ accuracy and $0.875$ pass@5), while compact quantized models such as SmolLM2-135M-Q4 enable efficient deployment with sub-$1$~GB memory and high throughput ($>760$ tokens/s), albeit with reduced reasoning fidelity. These findings demonstrate that quantization is a model-dependent, system-level design tradeoff and that no single model can simultaneously satisfy all performance constraints, implying the need for a hierarchical, heterogeneous multi-agent paradigm in 6G. Overall, this work establishes both a conceptual and empirical foundation for Agentic AI-Native 6G, highlighting key cross-layer challenges and positioning future networks as semantic, self-reasoning infrastructures capable of autonomous, policy-constrained adaptation.

\bibliographystyle{IEEEtran}
\bibliography{bibliography}

\end{document}